\documentclass[11pt, a4paper]{article}
\usepackage{natbib}
\usepackage[english]{babel}
\usepackage[utf8]{inputenc}
\usepackage[T1]{fontenc}
\usepackage[left=2.5cm,right=2.5cm,top=3cm,bottom=3cm]{geometry}
\usepackage{amsmath}
\usepackage{amsfonts}
\usepackage{amssymb}
\usepackage{dsfont} 
\usepackage{xcolor}
\usepackage[normalem]{ulem}

\usepackage{graphicx}
\usepackage{blindtext}
\usepackage{multicol} 
\DeclareUnicodeCharacter{02B9}{'}
\usepackage{fancyhdr} 
\numberwithin{equation}{section}
\usepackage{authblk}

\newcommand{\para}{\parallel}
\newcommand{\br}{\mathbf{r}}

\begin{document}
\title{\textbf{Inference of the 3D pressure field exerted by a single cell from a thin membrane transverse deformation}}
\author{Quentin Bédel$^{1,2}$ Lo\"ic Dupr\'e$^{2,3}$, Nicolas Destainville$^{1}$} 
\affil{1: Univ Toulouse, CNRS, LPT, Toulouse, France \\ 
       2: Univ Toulouse, Inserm, CNRS, INFINITy, Toulouse, France \\
       3: Department of Dermatology, Medical University of Vienna, Austria
       }

\maketitle
\begin{abstract}
	Numerous cell types relate to their immediate environment by exerting a three-dimensional pressure field on their environment, with components both longitudinal and transverse to the cell membrane. This pressure field can in principle be measured by traction force microscopy experiments. Compared to other approaches, the technique of Protrusion Force Microscopy gives access with high spatial resolution to the pressure field by measuring the deformation of a thin elastic membrane using atomic force microscopy (AFM). However, while the pressure field under interest is three-dimensional, the height profile measured by AFM is only one-dimensional. We propose a solution to this inverse problem and we explore its regime of applicability in the experimental context. 
\end{abstract}

\section{Introduction}

Mechanosensing is the capacity of a cell to probe the mechanical stimuli of its environment and to orient consequently its decisions concerning key processes such as apoptosis, migration, differentiation or morphogenesis~\citep{Chen2017}. Reciprocally, multiple cells also exert forces on their environment
to execute their biological function, among which immune cells. Interestingly, lymphocytes, including T cells, are considered as poorly adherent when they patrol the different tissues of the organism, but switch to a very adherent behavior when they encounter the antigen for which they are specific. Indeed a stable immunological synapse is then established between the T-cell and its cellular target (virus-infected cell or tumor cell).

The forces acting within this immunological synapse play a crucial role in various biomechanical processes~\citep{Dillard2014,Basu2016,Sawicka2017,Basu2017,Roy2018,Wahl2019,Tamzalit2019,Blumenthal2020,Huse2025}. A deeper understanding of their molecular origins at the level of the active cytoskeleton network, as well as the local quantification of their magnitude and direction, enable the identification of the influence of forces on different synaptic mechanisms but remain challenging from an experimental viewpoint. The potential implications embrace the development of therapeutics targeting cancers and inflammatory or autoimmune diseases. 

\begin{figure}
	\centering
    \includegraphics[width=9cm]{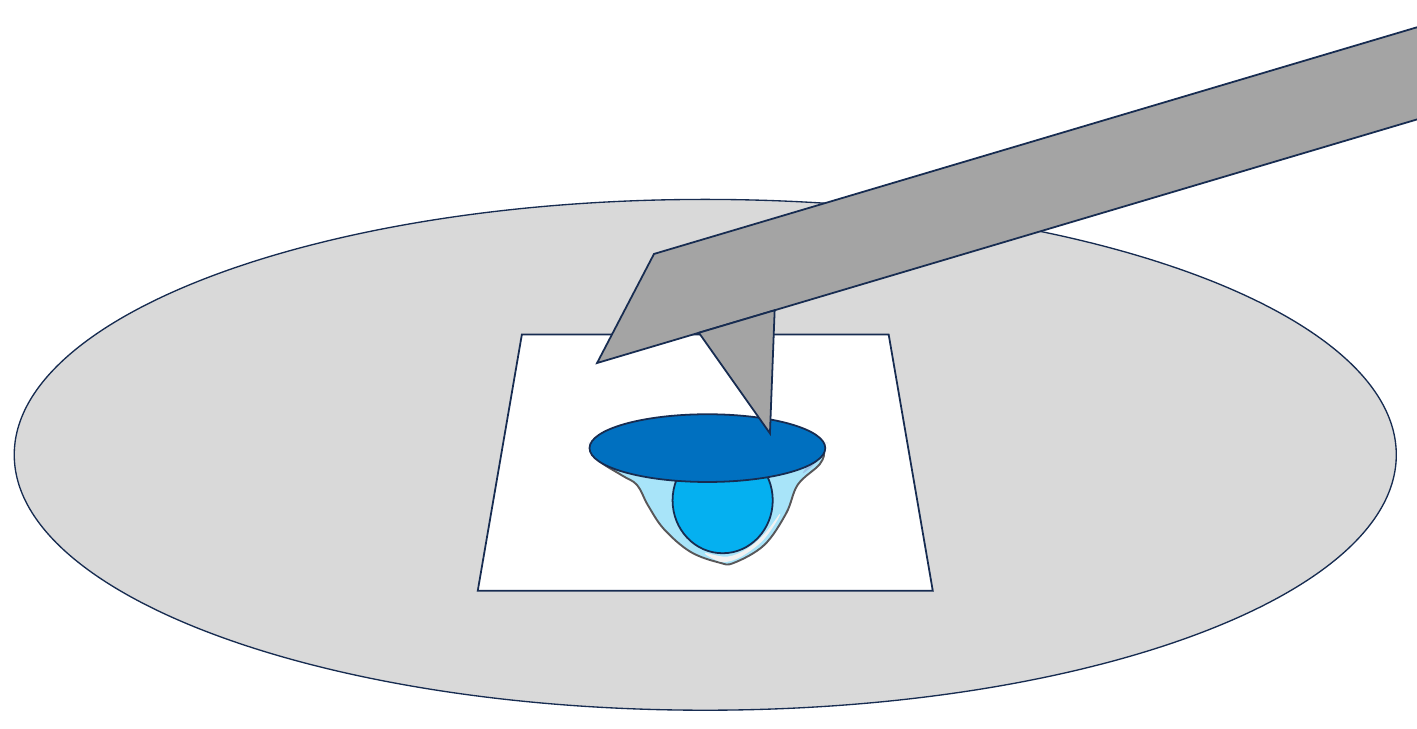} \\ \ \\
        \includegraphics[width=9.5cm]{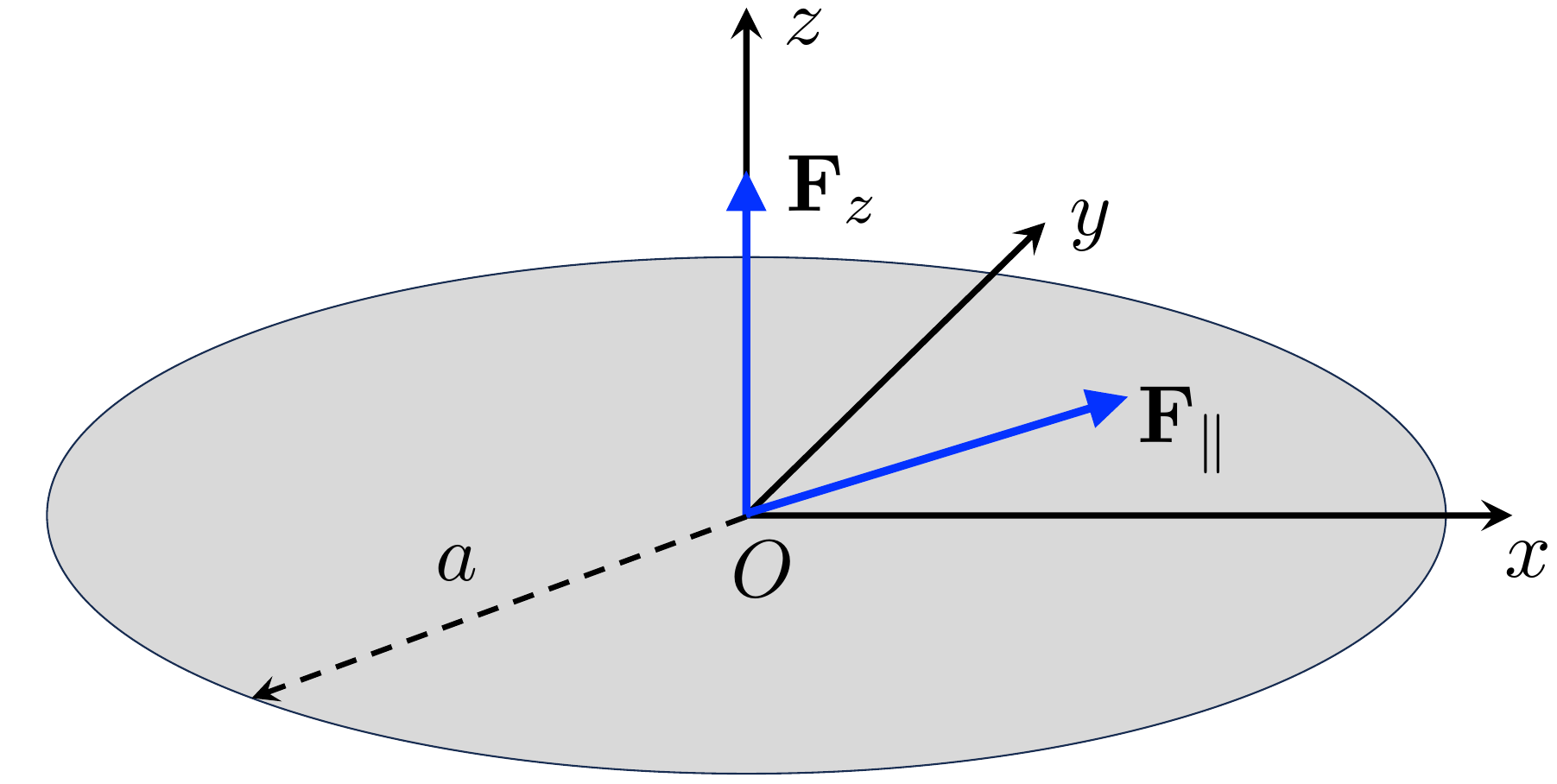}
    \caption{Top: Cartoon of the Protrusion Force Microscopy experimental setup (not at scale). The cell (in blue, its interface with the stimulatory substrate in dark blue) adheres on the lower face of the elastic membrane (in light gray), whereas the AFM measurement are performed on its upper face (AFM canterliver in dark gray). The Region of Interest (ROI) scanned by the AFM appears in white. Bottom: Point loading of the circular elastic membrane of radius $a$ at its center $O$. $\mathbf{F}_z$ and $\mathbf{F}_\parallel$ are respectively transverse and longitudinal point loadings.
    \label{cartoon}}
\end{figure}

Over the past decades, several experimental techniques based on fluorescence microscopy have been developed to measure forces at the subcellular scale, in the nano-Newton range. Notable examples include the measurement of longitudinal forces via elastic gel deformation and micropillar bending~\citep{butler2002traction,Ghibaudo2008,Bashour2014,schwarz2015traction,mandal2023wasp}. 
Alternatively, measurements of cellular-induced deformation of a Formvar elastic membrane using atomic force microscopy (AFM) have been employed to quantify the transverse forces exerted by macrophages~\citep{labernadie2014protrusion}. This Protrusion Force Microscopy technique is schematized in Figure~\ref{cartoon} (Top). However, a limitation of this latter method is that only forces perpendicular to the substrate have been primarily considered so far, as they are inferred from the measured height of the Formvar deformation by the help of elasticity theory, whereas transverse and longitudinal forces can have the same order of magnitude~\citep{Aramesh2021}. In this study, we extend the analytical framework describing the deformation height of a loaded membrane to account for both transverse and longitudinal deformations under a three-dimensional pressure field. We then propose a mathematical model that enables the reconstruction of this three-dimensional pressure field components solely from the  measurement of the deformation scalar height field, relying on the resolution of an optimisation problem.

The paper is organized as follows. In the next section, using the theory of elasticity of plates, we calculate the deformation of the membrane under loading, for a transverse pressure field, a longitudinal one, and a combination of both. In section 3, we show how to solve the inverse problem of inferring the 3D pressure field from the measurement of the scalar height field of the membrane and we illustrate our findings on realistic examples related to T-cell synapses, before concluding in the last section.

\section{Deformation of a membrane under loading}

We model the elastic membrane as a circular elastic plate of radius $a$ and thickness $e$, with fixed boundary conditions as detailed below. Using elastic theory at the linear level, we first compute the deformation of the membrane under various point forces and pressure fields.  These results will be useful below to solve the inverse problem. We emphasize that, in both continuous and discrete descriptions, since the 3D loading field has the dimension of pressure (Pa), we express it accordingly rather than as a ``traction force'', and we adopt this convention throughout the paper.

\subsection{Deflection of a circular membrane under tension by a transverse point loading}
\label{transverse:prop}

The deflection along the $z$ direction of a circular clamped membrane of radius $a$, under residual tension $\tau > 0$ for a transverse point loading $P_z = F_z\delta(\mathbf r)$ at its center $O$ at $\mathbf r = \mathbf 0$, parallel to the membrane axis of symmetry (Figure~\ref{cartoon}, Bottom), satisfies, for small gradients:

\begin{align}
	u_z (r) &= \frac{F_za^2}{2\pi  \kappa k^2} \left\{  \frac{(K_1(k)-\frac{1}{k})}{I_1(k)}\left(I_0(k) - I_0 \left( \frac{kr}{a}\right) \right) + \left(K_0(k) - K_0 \left( \frac{kr}{a} \right) \right) - \text{ln} \left( \frac{r}{a} \right)  \right\}
	\label{u_z}
\end{align} where $I_{n}$ and $K_{n}$ are respectively modified Bessel functions of first and second kind of order $n$~\citep{abramowitz1968handbook}; $u_z(\mathbf r)$ is the deflection of the membrane, $r = \|\mathbf r\| = \sqrt{x^2+y^2}$ is the distance to the center $O$, 
$\kappa =\frac{E e^3}{12(1 - \nu^2)} $ is the bending rigidity with $E$ the Young modulus, $e$ the membrane thickness, $\nu$ the Poisson ratio~\citep{landau2012theory} and $k = a\sqrt{\frac{ \tau}{\kappa}}$ is the dimensionless inverse Helfrich correlation length~\citep{hong1990measuring}.
The calculation of this transverse deformation is detailed in appendix \ref{app0}.

In the following, we will use these results beyond the small-gradient regime for the sake of simplicity and discuss the implications of this in the last section. 

\subsection{Longitudinal deformation of a circular membrane under longitudinal point loading}

Now we examine the membrane longitudinal deformation under longitudinal loading $\mathbf P_{\parallel}(\mathbf{r}) = P_x (\mathbf{r}) \mathbf e_x + P_y (\mathbf{r}) \mathbf e_y $. On an infinite tensionless membrane ($a \rightarrow \infty$), it is solution of the coupled partial differential equations~\citep{landau2012theory}

\begin{equation}
E e \left[ \frac{1}{1-\nu^2} \frac{\partial^2 u_x}{\partial x^2} +  \frac{1}{2(1+\nu)} \frac{\partial^2 u_x}{\partial y^2}+ \frac{1}{2(1-\nu)} \frac{\partial^2 u_y}{\partial y \partial x} \right]= - P_x 
	\label{eq_longitudinale_Landau_x}
\end{equation}
\begin{equation}
		E e \left[ \frac{1}{1-\nu^2} \frac{\partial^2 u_y}{\partial y^2} +  \frac{1}{2(1+\nu)} \frac{\partial^2 u_y}{\partial x^2}+ \frac{1}{2(1-\nu)} \frac{\partial^2 u_x}{\partial x \partial y} \right]= - P_y 
	\label{eq_longitudinale_Landau_y}
\end{equation}
In thin plate theory, the longitudinal deformation is assumed to be homogeneous across the entire plate (i.e. independent of the coordinate $z$), since the applied forces are considered small as compared to the internal constraints, as demonstrated in~\citep{landau2012theory}. Therefore, they can be treated as mid-plate forces, and by up-down symmetry  $z\rightarrow -z$, we have $u_z = -u_z$ for the same longitudinal loading, thus $u_z = 0$. Longitudinal loading of the membrane does not cause any vertical displacement.


For a point loading $\mathbf{P}(\mathbf{r})=\mathbf{F}_\parallel \delta (\mathbf{r})$ (Figure~\ref{cartoon}, Bottom), one can define the characteristic lengths $\tilde f_\alpha = F_{\parallel,\alpha}/(Ee)$, $ \alpha \in \{x,y\}$, which set the typical length-scale of the deformation caused by the loading (see below).
Considering first the limit $a \rightarrow \infty$, we Fourier-transform the previous displacement equations, and obtain: 
\begin{align}
	 \frac{1}{1-\nu^2} q_x^2 \hat{u}_x + \frac{1}{2(1+\nu)} q_y^2 \hat{u}_x +  \frac{1}{2(1-\nu)} q_y q_x\hat{u}_y &=  \tilde f_x \\
	 \frac{1}{1-\nu^2}  q_y^2 \hat{u}_y +  \frac{1}{2(1+\nu)} q_y^2 \hat{u}_x + \frac{1}{2(1-\nu)} q_y q_x\hat{u}_x &=\tilde f_y
\end{align} where $\mathbf{q} = (q_x,q_y)$ is the wave-vector. 
Thus: 
\begin{align}
\hat u_x &= \frac{\tilde{f}_x[ (1-\nu^2) q^2_x + 2(1+\nu) q^2_y]  - \tilde{f}_y (1+\nu)^2 q_x q_y}{q^4}\\
\hat u_y &=  \frac{\tilde{f}_y[ (1-\nu^2) q^2_y + 2(1+\nu) q^2_x]  - \tilde{f}_x (1+\nu)^2 q_y q_x}{q^4}
\end{align} 
where $q = \|\mathbf{q}\|$. Consequently, if $\mathcal{F}^{-1}$ denotes the inverse Fourier transform, we can write

\begin{equation}
	u_x =  -\tilde f_x \left( (1-\nu^2) \frac{\partial^2}{\partial x^2}\mathcal{F}^{-1}\left[\frac{1}{q^4}\right] +2(1+\nu) \frac{\partial^2 }{\partial y^2}\mathcal{F}^{-1}\left[\frac{1}{q^4}\right] \right) + \tilde{f}_y(1+\nu)^2   \frac{\partial^2 }{\partial y \partial x}\mathcal{F}^{-1}\left[\frac{1}{q^4}\right] 
	\label{ux_tf_inverse}
\end{equation}
\begin{equation}
u_y =  -\tilde f_y \left( 2(1+\nu)  \frac{\partial^2}{\partial x^2}\mathcal{F}^{-1}\left[\frac{1}{q^4}\right] + (1-\nu^2) \frac{\partial^2 }{\partial y^2}\mathcal{F}^{-1}\left[\frac{1}{q^4}\right] \right) + \tilde{f}_x (1+\nu)^2 \frac{\partial^2 }{\partial y \partial x}\mathcal{F}^{-1}\left[\frac{1}{q^4}\right]
	\label{uy_tf_inverse}
\end{equation}
The inverse $2$D Fourier transform of $\frac{1}{q^4}$ is not properly defined because of the divergence at $q=0$ but one can verify, by injecting Eqs.~\eqref{ux_tf_inverse} and \eqref{uy_tf_inverse} in Eqs.~\eqref{eq_longitudinale_Landau_x} and \eqref{eq_longitudinale_Landau_y}, that in fact $\mathcal{F}^{-1}\left[\frac{1}{q^4}\right]$ can be replaced by any Green function $\psi$ of the $2D$ bilaplacian, satisfying
\begin{equation}
\Delta \Delta \psi(\mathbf r) = \delta (\mathbf r)
\label{biharmonic_g}
\end{equation} where $\delta (\mathbf r)$ is the Dirac distribution.
The classic solution of this partial differential equation reads:
\begin{equation}
	\psi(\mathbf r) = \frac{r^2}{8\pi} \ln \left(\frac{r}{a}\right) + \phi(\mathbf r)
	\label{psi:phi}
\end{equation} where the first term on the right hand-side is a particular solution of Equation~\eqref{biharmonic_g} and $\phi(\mathbf r)$ is any biharmonic function, i.e. satisfying  $\Delta \Delta\phi(\mathbf r) = 0$, also known as a Michell solution when expressed in polar coordinates~\citep{michell1899direct}.

Considering for simplicity and without loss of generality at this stage, a force $\textbf{F}=F_x \textbf{e}_x$  along the $x$ axis, Eqs.~\eqref{ux_tf_inverse} and \eqref{uy_tf_inverse} become: 
\begin{eqnarray}
	u_x(\mathbf{r}) & =&  -\tilde f_x (1+\nu) \left( (1-\nu) \frac{\partial^2}{\partial x^2}\psi(\mathbf r) + 2 \frac{\partial^2 }{\partial y^2}\psi(\mathbf r) \right) \\
u_y(\mathbf{r}) &=& \tilde{f}_x (1+\nu)^2 \frac{\partial^2 }{\partial y \partial x}\psi(\mathbf r)
\end{eqnarray}

In the case of interest where the size $a$ of the system is finite, imposing the fixed-boundary condition where the in-plane displacement $\mathbf{u}_\parallel(a,\theta)=0$, and identifying the adequate Michell solution $\phi(r,\theta)$, we show in Appendix \ref{appA} that:
\begin{eqnarray}
	\mathbf{u}_{\parallel}(r,\theta) & = & -\frac{(\nu+1)^2(3\nu-1)}{8\pi(\nu-3)}\tilde{f}_x \left[\left(\frac{r}{a}\right)^2-1\right] \cos \theta \, \mathbf{e}_r \nonumber \\ 
	& + & \frac{(\nu+1)^2(\nu+5)}{8\pi(\nu-3)}\tilde{f}_x \left[\left(\frac{r}{a}\right)^2-1\right] \sin \theta \, \mathbf{e}_\theta  
	+ \frac{(\nu-3)(\nu+1)}{4\pi} \tilde{f}_x  \ln\left(\frac{r}{a}\right) \, \mathbf{e}_x  
	\label{u_r}
\end{eqnarray}
The displacement at the membrane boundary $r=a$ vanishes, as required. For any longitudinal point force $\mathbf{F}_\parallel$, now not necessarily parallel to $\textbf{e}_x$, this propagator can now be written
\begin{eqnarray}
	\mathbf{u}_{\parallel}(\mathbf{r}) & = & -\frac{(\nu+1)^2(3\nu-1)}{8\pi(\nu-3)} \left[\left(\frac{r}{a}\right)^2-1\right] \frac{\tilde{\mathbf{f}}_\parallel\cdot\mathbf{r}}{r^2} \mathbf{r} \nonumber \\ 
	& + & \frac{(\nu+1)^2(\nu+5)}{8\pi(\nu-3)} \left[\left(\frac{r}{a}\right)^2-1\right] \frac{\tilde{\mathbf{f}}_\parallel \times \mathbf{r}}{r^2} \times \mathbf{r}
	+ \frac{(\nu-3)(\nu+1)}{4\pi}   \ln\left(\frac{r}{a}\right) \tilde{\mathbf{f}}_\parallel  
	\label{u_r_vect}
\end{eqnarray}
where $\tilde{\mathbf{f}}_\parallel = \mathbf{F}_\parallel/(Ee)$. 

Those boundary conditions come from our assumption that the system is circular and clamped. For a rectangular system, the calculations would become more involved and we would have to appeal to the generalized Michell solution \citep{michell1899direct}, which goes beyond the scope of the present work. 

To finish with, note that to get rid of the non-physical logarithmic divergence at $r=0$, we can more realistically not take a point loading where $\mathbf{P}_{\parallel}(\mathbf{r}) = \mathbf{F}_{\parallel}\delta (\mathbf{r})$, but a homogeneous pressure field of intensity $F_\parallel/(\pi b^2)$ on a small disk $\mathcal{D}(O,b)$ of radius $b \ll a $, centered at the origin $O$. Considering again a force along $\mathbf{e}_x$ without loss of generality, the resulting displacement of the disk center at $\mathbf{r}=0$, also along $\mathbf{e}_x$ by symmetry, reads 
\begin{align}
    \mathbf{u}_{\parallel,b}(\mathbf{0}) &= \frac1{\pi b^2} \int_{\mathcal{D}(O,b)} {\rm d}^2\mathbf{r}_0 \, \mathbf{u}_{\parallel}(\mathbf{0}-\mathbf{r_0})\\
    &= \tilde{f}_x \left\{  \frac{(\nu+1)^2(2-\nu)}{4\pi(\nu-3)}  \left[\frac{b^2}{4a^2}-\frac{1}{2}\right] +   \frac{(\nu-3)(\nu+1)}{4\pi} \left[2\ln\left(\frac{b}{a}\right) - 1\right] \right\} \mathbf{e}_x
\label{regul}
\end{align}
where we have assumed that the propagator~\eqref{u_r_vect} is unchanged (at leading order) when the longitudinal force is applied at a point $\mathbf{r}_0$ just next to the origin, i.e. $\| \mathbf{r}_0 \| \leq b \ll a$ (see end of Section~\ref{pressure:field}).
Since $b\ll a$, the displacement of the origin is dominated by the last term in this expression.  It is thus proportional to $\ln (b/a)$, which regularizes the divergence. 

\subsection{Adding the uniform longitudinal tension}

We consider again a uniform residual frame tension $\tau$, for a $2D$ membrane. In absence of any loading, Hooke's law gives \citep{landau2012theory}: 
\begin{align}
	\sigma_{\alpha\beta} &= \frac{E}{1+\nu} \left(u^0_{\alpha\beta} + \frac{\nu}{1-2\nu} u^0_{\gamma\gamma}\delta_{\alpha\beta}\right) \equiv \tau 
	\label{sigma_tension}
\end{align} where $\sigma_{\alpha \beta}$ is the $\alpha,\beta \in \{x,y\}$ component of the stress tensor $\bar{\bar{\sigma}}$, $\delta_{\alpha\beta}$ is the Kronecker symbol, and $u^0_{\alpha\beta}$ is the  component of the deformation tensor, defined by:
\begin{equation}
	u^0_{\alpha \beta} = \frac{1}{2}\left(\frac{\partial u^0_\alpha}{\partial \beta} + \frac{\partial u^0_\beta}{\partial \alpha}\right)
	\label{uij_u}
\end{equation}
Because the tension is uniform within the membrane, we have $u_{xy}^0 =0$ and 
\begin{equation}
u^0_{xx} = u^0_{yy} = \frac{(1+\nu)(1-2\nu)}{E}\tau \equiv \tilde\tau 
\end{equation}
We finally obtain the isotropic in-plane displacement by integrating these expressions: 
\begin{align}
u^0_\alpha &= \tilde\tau \alpha
\end{align} 
that is to say $u^0(\mathbf{r})=\tilde\tau \mathbf{r}$, the expected uniform membrane deformation. 
Because we work at small deformations and we neglect the higher orders, we assume that considering at the same time a loading point and the uniform frame tension is equivalent to adding the loading point contribution to a membrane already stressed by tension. Then if $a$ is considered as the membrane radius when it is already under tension, as is the case in experiments, it eventually allows us not to explicitly take into account the frame tension in the longitudinal displacement~\eqref{u_r_vect}, contrary to the transverse loading above.

\subsection{Membrane under any pressure field P applied close to the membrane center}
\label{pressure:field}

For a $3D$ point force applied at the center of a circular membrane of radius $a$, with tension $\tau$, the total displacement field $\mathbf{u}(\mathbf{r})$ is thus the linear combination of longitudinal and vertical deflections in Eqs.~\eqref{u_z} and \eqref{u_r_vect} as: 
\begin{eqnarray}
	\mathbf{u}_{\parallel}(\mathbf{r}) & = & -\frac{(\nu+1)^2(3\nu-1)}{8\pi(\nu-3)} \left[\left(\frac{r}{a}\right)^2-1\right] \frac{\tilde{\mathbf{f}}_\parallel\cdot\mathbf{r}}{r^2} \mathbf{r} \nonumber \\ 
	& + & \frac{(\nu+1)^2(\nu+5)}{8\pi(\nu-3)} \left[\left(\frac{r}{a}\right)^2-1\right] \frac{\tilde{\mathbf{f}}_\parallel \times \mathbf{r}}{r^2} \times \mathbf{r}
	+ \frac{(\nu-3)(\nu+1)}{4\pi}   \ln\left(\frac{r}{a}\right) \tilde{\mathbf{f}}_\parallel  \nonumber \\
	u_z(\mathbf r) &= & \tilde{f}_z \frac{6a^2(1-\nu^2)}{\pi e^2 k^2}  \nonumber \\ 
	& \times & \left[\frac{(K_1(k)-\frac{1}{k})}{I_1(k)}\left(I_0(k) - I_0  \left( \frac{kr}{a}\right) \right) + K_0(k) - K_0 \left( \frac{kr}{a} \right)  - \text{ln} \left( \frac{r}{a} \right)  \right]
	\label{eq_displacement}
\end{eqnarray} 
where we recall that $k=a \sqrt{\frac{\tau}{\kappa}}$. In practice, $a^2/(e^2k^2) \gg 1$ (see below) so that $\mathbf{u}_{\parallel}$ will appear to be a subdominant correction to $u_z$. In tensor notation, these relations become
\begin{equation}
	u_\alpha = G_{\alpha \beta} \tilde{f}_{\beta}
\end{equation}
where we use the Einstein summation notation for the space coordinates, now with $\alpha,\beta \in \{x,y,z\}$, and we introduce the second-order propagator tensor $\bar{\bar{G}}$ defined by Eqs.~\eqref{eq_displacement}. 
Working at the linear order thus allows us to uncouple the vertical displacement, function of vertical loading only, and the longitudinal displacement, function of longitudinal loading only. 

In principle, these propagators have been calculated for a point loading at the exact center of the membrane. We show in Appendix \ref{appB} that at dominant order, they are still valid for any point force exerted close to the membrane center. 

For any pressure field $\mathbf{P}(\mathbf r)$ exerted close to the membrane center, on a region much smaller than its radius $a$, we can thus write: 
\begin{equation}
	u_{\alpha} =  G_{\alpha \beta} * {\tilde P}_{\beta}
\label{eq_displacement_propagator}
\end{equation} where $*$ is the convolution product between the propagator $G_{\alpha \beta}$ of the deformation, defined through Equations~\eqref{eq_displacement} and the 3D pressure field.

\section{Inverse problem: Finding the 3D pressure field by knowing only the height of the deformation}
In experimental AFM measurements, we have in general only access to the information on the one-dimensional height of the deformation $h(\mathbf r)$~\citep{labernadie2014protrusion,poilane2000analysis,huse2020microparticle}, whereas the pressure field exerted by the cell on the substrate is 3-dimensional. Here, we propose a strategy to solve this apparent issue, inspired from Reference~\citep{huse2020microparticle}.

\begin{figure}
	\centering
    \includegraphics[width=9cm]{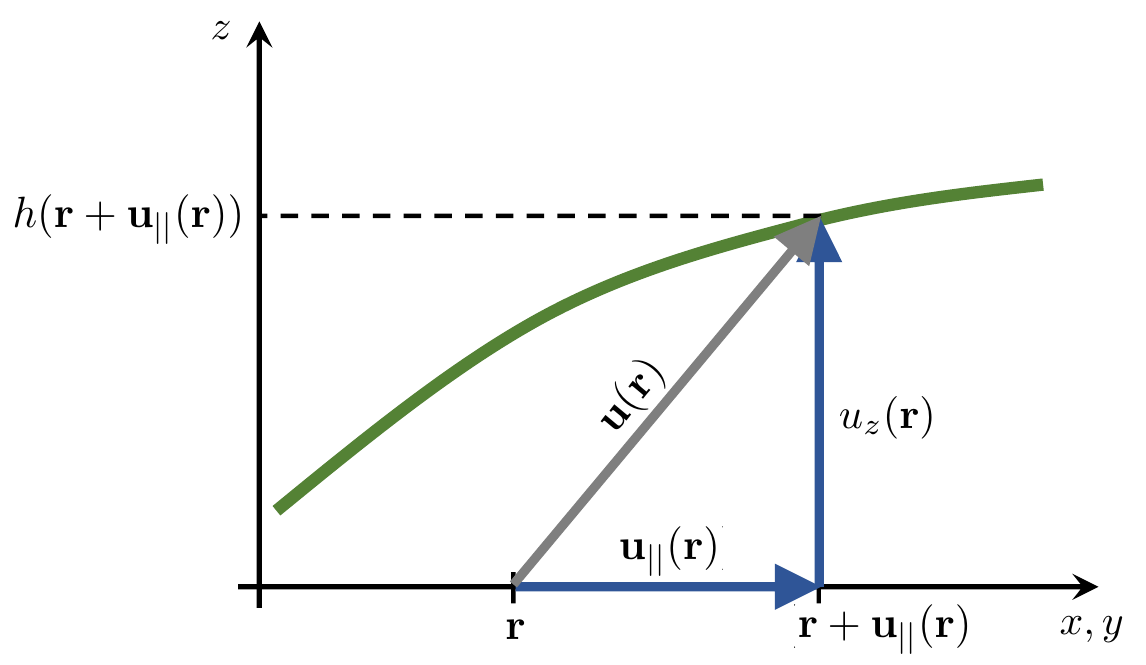}
    \caption{The height function $h$ (graph in green) results from the linear combination of the 2D longitudinal component $\mathbf{u}_\parallel$ and the transverse one $u_z$ (both in blue) of the total deformation field $\mathbf{u}$ at any point $\mathbf{r}$ of the membrane. This translates into Equation~\eqref{ZE:eq}.
    \label{h:vs:u}}
\end{figure}

For any displacement field $\mathbf{u}_{\parallel}$, the height field $h$ satisfies
\begin{equation}
h(\mathbf{r} + \mathbf{u}_\parallel(\mathbf r)) = u_z(\mathbf r)
\label{ZE:eq}
\end{equation}
everywhere, as illustrated in Figure~\ref{h:vs:u}.  We use the first-order approximation
\begin{align}
h(\mathbf r) \simeq u_z (\mathbf r) - \mathbf{u}_{\parallel}(\mathbf r) \cdot \boldsymbol{\nabla}h(\mathbf r) 
\label{h_dl}
\end{align}

By injecting Equation~\eqref{eq_displacement_propagator} in Equation~\eqref{h_dl}, we get for any $\mathbf{r}$
\begin{align}
h(\mathbf r) &= u_z(\mathbf r) - u_x(\mathbf r) \frac{\partial h}{\partial x}(\mathbf r) - u_y(\mathbf r) \frac{\partial h}{\partial y}(\mathbf r) \label{h:vs:u2}\\
&= H_\alpha (\mathbf r) (G_{\alpha \beta} * P_\beta)(\mathbf{r})
\end{align} 
where we introduce the first-order tensor $\mathbf{H}(\mathbf{r}) = \left(-\frac{\partial h}{\partial x}(\mathbf r), -\frac{\partial h}{\partial y}(\mathbf r), 1\right)$.
We can finally write the measured deflection $h(\mathbf r)$ as a linear combination of the (unknown) $3D$ components of the pressure field $\mathbf P(\mathbf r)$, formally:
\begin{equation}
h(\mathbf r) = H_\alpha (\mathbf r)\int_{\mathbb{R}^2} {\rm d}^2\mathbf{r'} G_{\alpha \beta} (\mathbf{r}- \mathbf{r}')  P_\beta(\mathbf{r}') 
	\label{h_g}
\end{equation}

\subsection{Discretization}

In experiments, one only has access to a discrete information, the height function $h$ being measured on each pixel of index $k$ in the $2$D image. We discretize the Region of Interest (ROI, Figure~\ref{cartoon}) where the height field $h$ is measured on $N$ square pixels of side-length $ \ell$. We denote by $N_c<N$ the number of pixels where the cell dwells (called the ``cell support'' thereafter, and denoted by $\mathcal{S}$). Equation~\eqref{h_g} becomes:
\begin{align}
	h^k &=  H_\alpha^k \ell^2 \sum_{i}   G_{\alpha\beta}^{i,k} P^{i}_\beta 
	\label{h_discrete}
\end{align}
where the longitudinal propagator is regularized at $\mathbf{r}=\mathbf{0}$ according to Equation~\eqref{regul} with $b=\ell /2$. The discrete partial derivatives of $h$ appearing in the $H_\alpha^k$ are calculated from the measured height function $h$ using a symmetric derivative scheme involving two nearest neighbor pixels. 

Since the pressure field has to vanish outside the cell support, where there is no contact between the cell and the support and since cells can only apply forces at contact~\citep{Dembo1996}, we look in principle for solutions for the $3$ components of the pressure field for each of the $N_c$ support pixels, with $N$ linear constraints ensuing from the $N$ values of $h$ on each pixel of the whole system. 
There are thus 3 cases: 
\begin{itemize}
	\item $ N > 3N_c$:  There are more equations than unknowns and consequently no exact solution. 
	\item $ N = 3N_c$: There are as many unknowns as equations and consequently one solution. 
	\item $ N < 3N_c$: There are more unknowns than equations and consequently an infinite number of solutions forming an affine subspace of $\mathbb{R}^{3N_c}$.
\end{itemize}
Except in the marginal case $N = 3 N_c$, the problem is either overdetermined or underdetermined. Following Reference~\citep{huse2020microparticle}, we are led to adopt an alternative optimization strategy, where some function of the (discrete) pressure field $\mathbf{P}$ will be minimized, subject to the constraints~\eqref{h_discrete} ensuing from the measured height field, with the help of Lagrange multipliers.

\subsection{Optimization problem}

We first determine the function that will be minimized, sum of two terms. First, since we cannot impose anymore that forces vanish outside the cell support $\mathcal{S}$, we seek to minimize the quadratic force excess outside $\mathcal{S}$~\citep{huse2020microparticle}
\begin{equation}
	 R[\mathbf{P}]=\sum_{i \in \mathrm{ROI}\setminus\mathcal{S}} \|\mathbf{P}^i\|^2
	\label{min_p}
\end{equation} 
where ROI still denotes the Region of Interest where the displacement field $h$ is measured. 

In addition, as also discussed in Reference~\citep{huse2020microparticle}, the stability of the numerical method requires to use a low-pass filter to avoid rapid spatial fluctuations of the inferred pressure field exerted by the cell. We therefore seek to minimize concomitantly the second quadratic form
\begin{equation}
Q[\mathbf{P}]=\sum_{\langle i,j \rangle \in \mathcal{S}} \| \mathbf{P}^i - \mathbf{P}^j \|^2 
\label{Q:def}
\end{equation}
for nearest-neighbor pixels in $\mathcal{S}$, which enforces the pressure field to vary smoothly. $R$ and $Q$ cannot be minimized independently because the support and non-support forces are linked through the constraints. This is why we are led to combine them in a unique cost function~\citep{huse2020microparticle} $\Gamma[\mathbf{P}] = R[\mathbf{P}] + w Q[\mathbf{P}]$. The weight $w$, setting the relative contributions of both quadratic forms in the cost function, will be determined empirically below. 

The minimization is performed under two kinds of linear constraints. Firstly, as stated in Equation~\eqref{h_discrete}, the measured height field $h$ must coincide everywhere with the one induced by the pressure field: in the discrete formulation, for any pixel $k$ of the ROI, 
\begin{equation}
h^k =  H_\alpha^k \ell^2  \sum_i   G_{\alpha\beta}^{i,k} P^i_\beta  \equiv C_0[\mathbf{P},k]
\label{h_g2}
\end{equation}

Secondly, in the specific case of an immobile cell adhering to the substrate without any contact with the external world except through the elastic membrane~-- but not for a migrating cell experiencing viscous drag~--, ignoring gravity forces (in the pN range for a cell in water), the total force that it exerts on the substrate must vanish~\citep{Dembo1996}:
\begin{equation}
\mathbf{0} = \sum_{i \in \mathcal{S}} \mathbf{P}_i \equiv \mathbf{C}_1[\mathbf{P}]
\label{Lagrange:mu}
\end{equation}
Note that in this case, the cell generically acts in first approximation as a force dipole, except for an idealized system possessing specific symmetries. 

The function of the pressure field to be minimized is
\begin{equation}
F[\mathbf P] = \Gamma[\mathbf{P}] - \sum_{k \in \mathrm{ROI}} \lambda_k C_0[\mathbf{P},k] - \boldsymbol{\mu} \cdot \mathbf{C}_1 [\mathbf{P}]
\end{equation}
where we have introduced the scalar Lagrange multipliers $\lambda_k$ enforcing the constraints~\eqref{h_g2} on each pixel $k$ of the ROI and the 3D one $\boldsymbol{\mu}$ enforcing 
 the zero-force condition~\eqref{Lagrange:mu}.

Deriving $F[\mathbf P]$ with respect to the pressure field components and the Lagrange multipliers, one obtains the following linear system of $4N+3$ equations with $4N+ 3$ unknowns : $3N$ values of $P_\alpha^k$, $N$ values of $\lambda$ and the three coordinates of $\boldsymbol{\mu}$. 

\begin{align}
	 \left.\begin{array}{r}
	2P_\beta^i \text{ if }  i\notin \mathcal{S} \\
	2\left(\nu_i P_\beta^i  - \sum_{j\in V_i} P_\beta^j \right) \text{ if } i \in \mathcal{S}
	\end{array} \right\}
	 & =  H_\alpha \ell^2\sum_{k}\lambda_k G_{\alpha \beta}^{i,k} + \mu_\beta\\
	h^k & =  H_\alpha \ell^2\sum_i G_{\alpha \beta}^{i,k}  P^i_\beta \\
	\sum_i P_\alpha^i & = 0 \text{ for an immobile cell} 
\end{align}
where $V_i$ is the set of neighbors in $\mathcal{S}$ of the pixel $i$ and $\nu_i$ their number. Solving this linear system solves the inverse problem. 
The resulting $(4N+3)\times(4N+3)$ linear system ($15879 \times 15879$ in the examples below) is solved numerically using the {\tt linalg.solve} routine from the NumPy linear algebra module, which relies on the LAPACK routine {\tt gesv} implementing an LU decomposition with partial pivoting~\citep{NumericalRecipes}. 

\subsection{Illustrative examples}
\label{illustrative:ex}

Now we present some practical examples in the T-cell synapse context to illustrate the above strategy and test its range of validity. To quantify its ability to solve the inverse problem, we apply a realistic, known 3D pressure field to deform the membrane by first applying the in-plane force at the center of each pixel and then applying the out-of-plane force at the resulting displaced position to determine the deformation height, in accordance with Eq. \eqref{ZE:eq}. We then solve the inverse problem to reconstruct the pressure field in order to compare it to the initially applied one. 

Since we will realize below that the most delicate part of the inverse problem is the reconstruction of the longitudinal field, we introduce the two following quantities in order to compare the longitudinal applied and reconstructed fields on the cell support $\mathcal{S}$, respectively denoted by $\mathbf{P}^{\rm app}_\parallel$ and $\mathbf{P}^{\rm rec}_\parallel$, both vectors of $\mathbb{R}^{2N_c}$: the ratio of their Euclidean norms in $\mathbb{R}^{2N_c}$ and the scalar product of these vectors after being normalized:
\begin{eqnarray}
\rho & = & \frac{\| \mathbf{P}^{\rm rec}_\parallel \|}{\| \mathbf{P}^{\rm app}_\parallel \|}  \\
c & = & \frac{\mathbf{P}^{\rm app}_\parallel}{\| \mathbf{P}^{\rm app}_\parallel \|}  \cdot \frac{\mathbf{P}^{\rm rec}_\parallel}{\| \mathbf{P}^{\rm rec}_\parallel \|}
\end{eqnarray}
Here, the vectors $\mathbf{P}_\parallel$ are understood as vectors the coordinates of which are given by the $2N_c$ coordinates of the 2-dimensional vectors $\mathbf{P}^i_\parallel$ on the $N_c$ support pixels. 

If we were able to exactly solve the inverse problem, we would find $\rho=c=1$. Conversely, in Appendix~\ref{scalar:prod}, we demonstrate that if the reconstructed vector was random, we would find $c\sim 1/\sqrt{2N_c} \ll 1$.

In the following examples, we assume that the cell, of radius $R=5$~$\mu$m (the typical size of a T-cell~\citep{Hui2015}), is situated close to the center of the elastic membrane (Figure~\ref{cartoon}). The ROI, chosen a square of side $L=15$~$\mu$m, is much smaller than the elastic membrane of diameter $2a=100$~$\mu$m. We use a discretization of the ROI by $N=63 \times 63$ pixels of size $\ell=L/63\simeq 0.24$~$\mu$m. The other model parameters are given in Table~\ref{tab:parametres}. 

\begin{table}[h]
\centering
\begin{tabular}{|c|c|c|}
\hline
Parameter & Description & Value  \\ \hline \hline
$2a$ & Size of the membrane$^\dagger$ & 100~$\mu$m  \\ \hline
$E$ & Young's Modulus$^\dagger$ & 2.3~GPa  \\ \hline
$e$ & Membrane thickness & 5~nm  \\ \hline
$\tau_0$ & Residual bulk tension$^\ddagger$ & 100~Pa   \\ \hline
$\tau = \tau_0 e$ & Residual surface tension & $0.5$~$\mu$N/m \\ \hline
$\nu$ & Poisson's ratio$^\dagger$ & $0.33$  \\ \hline
$\kappa = \frac{Ee^3}{12(1-\nu^2)}$ & Membrane bending modulus & $2.7 \times 10^{-17} J$   \\ \hline
$k = a\sqrt{\frac{\tau}{\kappa}}$ & Inverse correlation length & $6.8$  \\ \hline
$F_{z, {\rm tot}}$  $F_{\parallel, {\rm tot}}$& Force components$^\ddagger$ & 10~nN  \\ \hline
\end{tabular}
\caption{Model parameters. $\dagger$: values from Reference~\citep{labernadie2014protrusion} for AFM experiments on a Formvar membrane. Formvar membranes of thickness 5~nm are commercially available. $\ddagger$: see text. }
\label{tab:parametres}
\end{table}

\subsubsection{Ideal T-cell synapse}

\begin{figure}
	\centering
    \includegraphics[height=5.45cm]{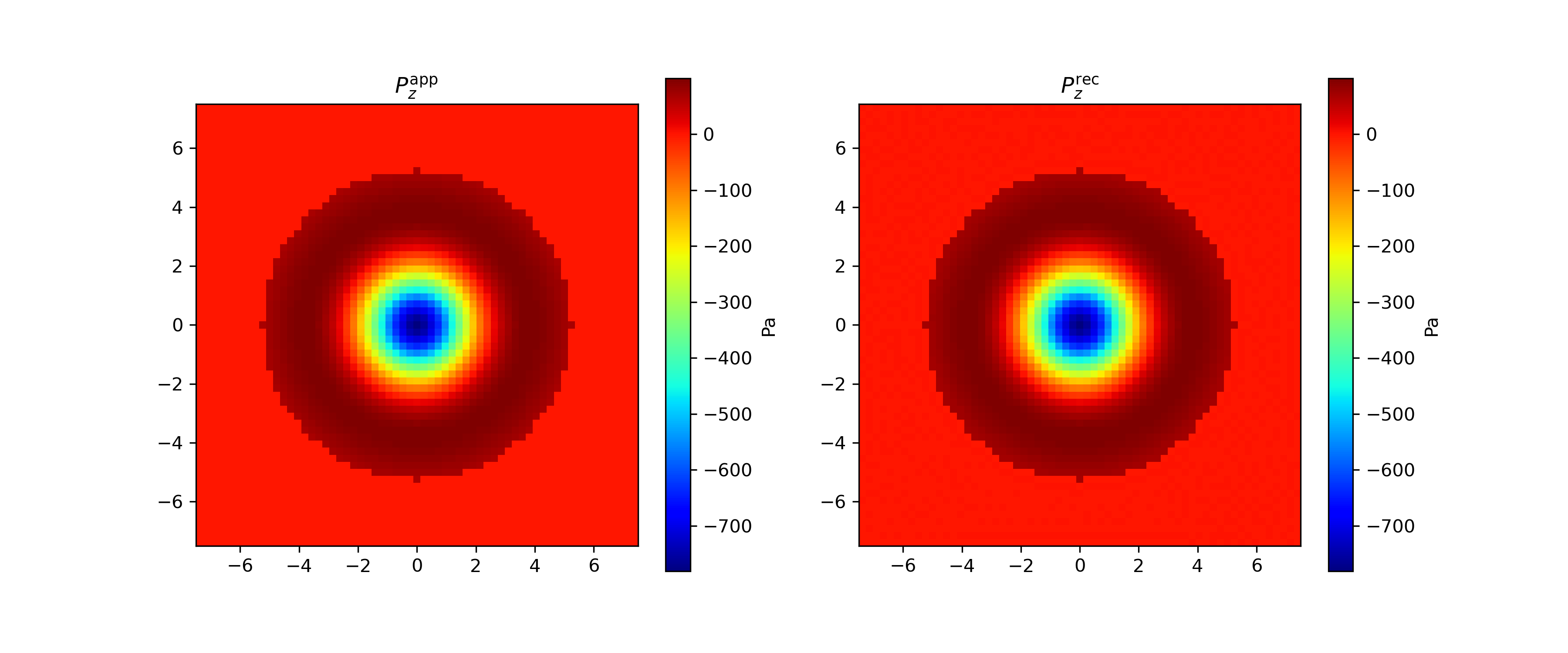} \\
    \includegraphics[height=5cm]{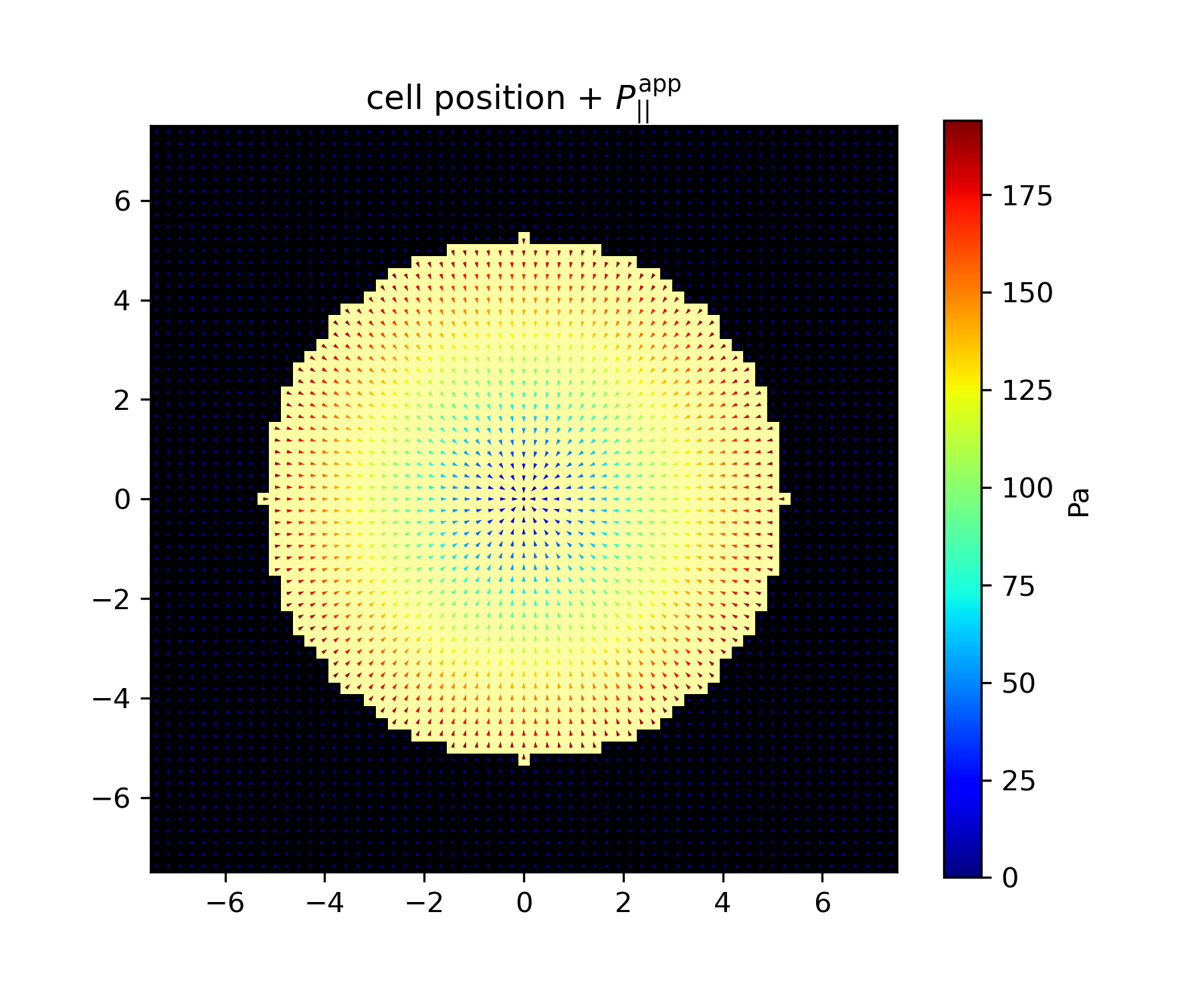}  \hspace{-0.8cm} \includegraphics[height=5cm]{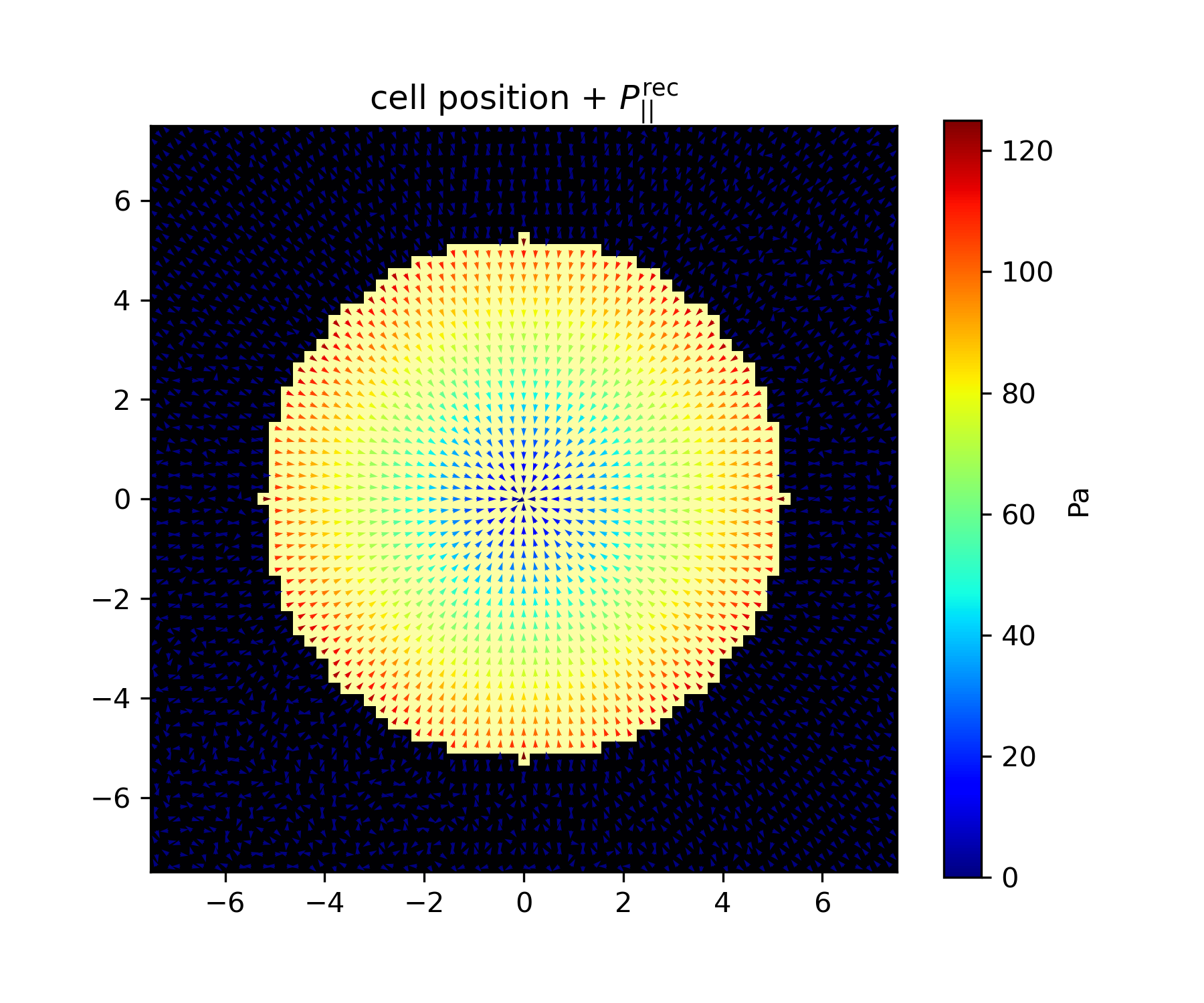}
    \caption{Ideal T-cell synapse. Pressure field $\mathbf{P}$ exerted by the cell at the synapse. Top-left: applied transverse pressure field $P_z(r)$  with $F_{z,{\rm tot}} =10$~nN. Top-right: reconstructed transverse pressure field. 
    Bottom-left: applied longitudinal pressure field $P_\parallel(r) \mathbf{e}_r$ with $F_{\parallel,{\rm tot}} =10$~nN. Bottom-right: reconstructed longitudinal pressure field. In the bottom figures, the direction of the force is indicated by the arrow and its intensity by the color map. The square ROI side length is $L=15$~$\mu$m. It is discretized in $N=63 \times 63$ pixels. The T-cell synapse radius is $R=5$~$\mu$m. The coordinates on the axes are given in $\mu$m. The color maps give the local pressure in Pa.
    \label{ideal:synapse}}
\end{figure}

We begin with an ideal T-cell synapse with axial symmetry. Its boundary is circular and the pressure field has the form $\mathbf{P}=P_z(r) \mathbf{e}_z + P_\parallel(r) \mathbf{e}_r$  inside the immunological synapse, as illustrated in Figure~\ref{ideal:synapse}. The functional forms of $P_z(r)$ and $P_\parallel(r)$ are chosen as follows. First, the net force must vanish for an immobile cell, as explained above. This is satisfied for any radial longitudinal pressure field $P_\parallel(r) \mathbf{e}_r$. As for the transverse one, we know from experiments~\citep{huse2020microparticle} that the T-cell pushes in the center of the immunological synapse. Consequently, it pulls in its periphery so that the total force vanishes. The chosen functional form to mimic these features is $P_z(r) = -A_1 e^{-\frac{r^2}{2\sigma_1^2}}+A_2 e^{-\frac{(r-r_2)^2}{2\sigma_2^2}}$ for $r\leq R$ (and $P_z(r) = 0$ if $r>R$). 

The longitudinal pressure field $P_\parallel(r)$ ensues from the centripetal cellular actin flow close to the T-cell synapse~\citep{Basu2017}. The actin flow drags the integrins ensuring the cohesion of the synapse, which results in a friction force exerted on the substrate locally proportional to the actin velocity in the linear regime~\citep{Sens2013}. The centripetal actin flow in this axisymmetric case being radial and its velocity having been predicted theoretically~\citep{CallanJones2008} and measured experimentally~\citep{Comrie2015} to be proportional to $r$, we therefore assume that $P_\parallel(r) = -P_0 r$ for $r\leq R$, in qualitative agreement with experimental data~\citep{Hui2015}.

The above non-negative constants $A_1$, $A_2$, $\sigma_1$, $\sigma_2$, $r_2$ and $P_0$ characterizing the pressure field are set by its total intensity, that we define as the integral of the norm of the pressure field exerted by the cell on its support:
\begin{eqnarray}
F_{z,{\rm tot}}  &=& \int_{\mathcal{S}}  {\rm d}^2\mathbf{r} \, | P_z (\mathbf{r}) | = \ell^2 \sum_{i \in \mathcal{S}} | P_z^i |  \\
F_{\parallel,{\rm tot}}  &=& \int_{\mathcal{S}}  {\rm d}^2\mathbf{r} \, \| \mathbf{P}_\parallel (\mathbf{r}) \| = \ell^2  \sum_{i \in \mathcal{S}} \| \mathbf{P}_\parallel^i \| 
\end{eqnarray}
expressed here both in the continuous and discrete versions. Below we will use the realistic reference value $F_{\parallel, {\rm tot}}=10$~nN~\citep{Basu2016,Tamzalit2019}, corresponding to an average pressure of 10 nN$/ \pi R^2 \sim 100$~Pa, a typical longitudinal pressure exerted by a T-cell on a rigid substrate~\citep{mandal2023wasp}. In addition, it has been measured by an alternative traction force microscopy technique relying on dispersed marker beads that local longitudinal and transverse forces are comparable in Jurkat T-cells~\citep{Aramesh2021}. Therefore we also use the value $F_{z,{\rm tot}} =10$~nN.
This led us to the following choice of parameter values:  $A_1=50$, $A_2=7.76$, $\sigma_1=R/4$, $\sigma_2=R/8$, $r_2=R/5$ and $P_0=0.5$.

To calibrate the value of the weight $w$ setting the relative contributions of the two quadratic forms in the cost function $\Gamma[h]$, we look for the value that optimizes the compromise between the correct orientation and magnitude of the reconstructed forces. Once this value $w=200$ is set empirically for the ideal synapse~\cite{Aramesh2021}, we keep it for the more realistic ones below. It depends on the model parameters.  

As illustrated in Figure~\ref{ideal:synapse}, the reconstructed pressure field reproduces correctly the applied one. The dominant source of the membrane deformation field $h$ being the transverse pressure $P_z$, the reconstructed transverse field reproduces very well the applied one (Figure~\ref{ideal:synapse}, top row). As far as the longitudinal pressure field is concerned, one recovers its axial symmetry as well the proportionality of $P_\parallel$ to $r$. Consistently, we find the very good value $c \simeq0.9995$. In contrast, we find $\rho \simeq 0.58$, meaning that the magnitude of the force field is about 58~\% of the applied one, which reflects that the transverse field being a first-order correction to $h$, it is less efficaciously reconstructed. We shall return to this point in the Discussion and propose a practical way to compensate for this effect.

\subsubsection{Robustness with respect to AFM noise}

\begin{figure}
	\centering
    \includegraphics[height=5cm]{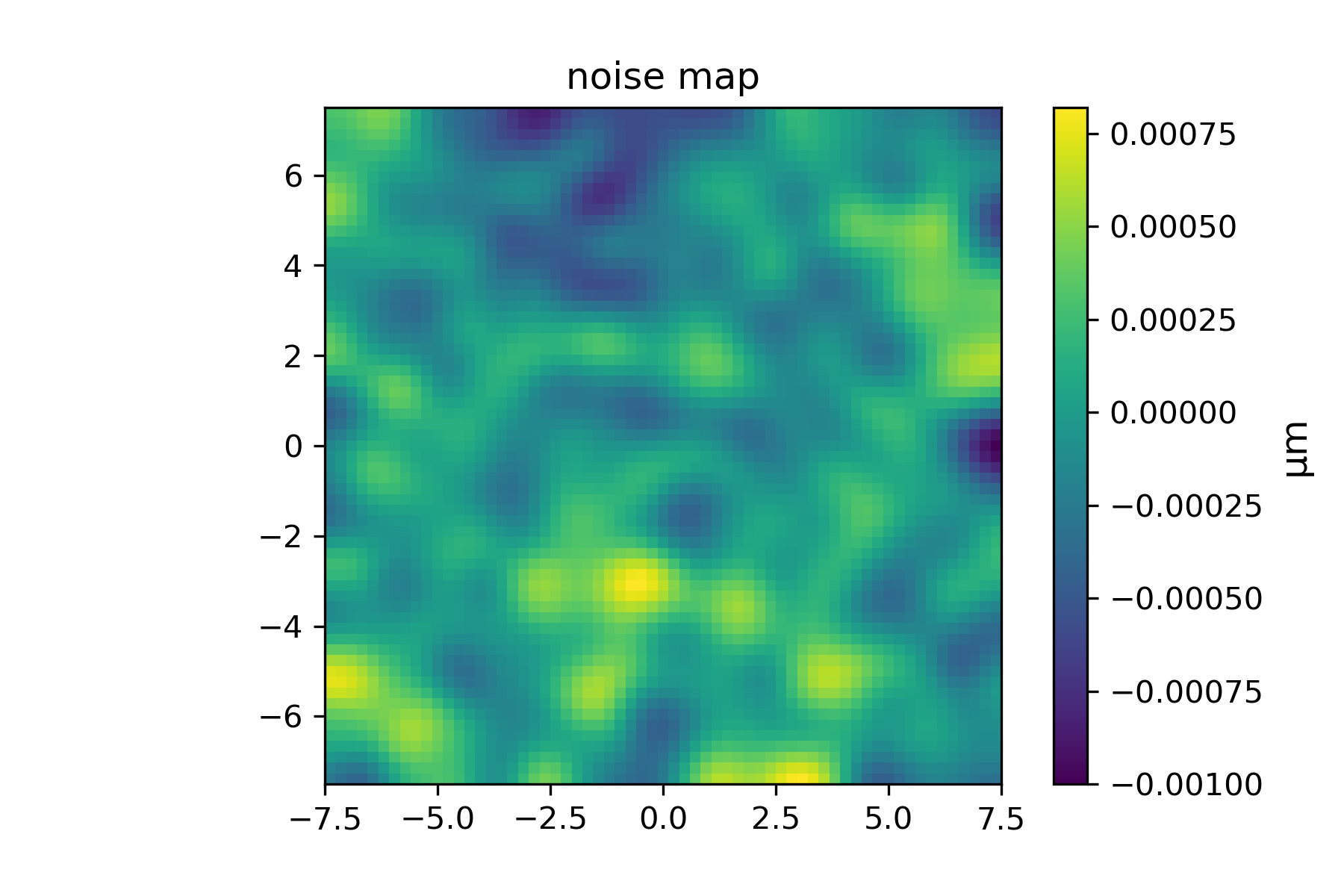}~\hspace{-0.8cm}~\includegraphics[height=5cm]{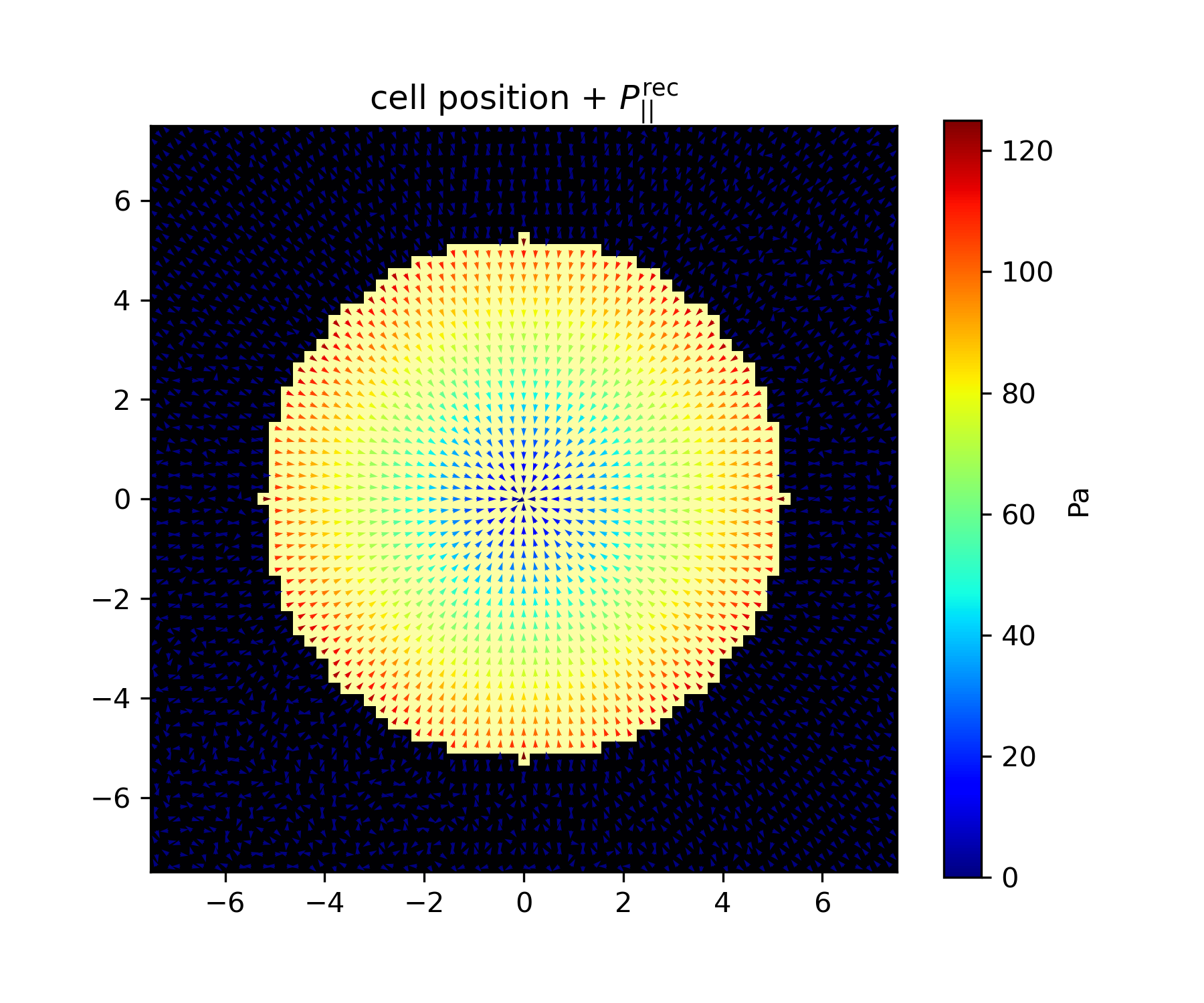}
    \caption{Left: Example of correlated AFM noise map. It is added to the height field $h$ before solving the inverse problem. Right: longitudinal pressure field reconstructed from the new height field. The applied pressure field is the same as in Figure~\ref{ideal:synapse}. All distances are in $\mu$m. 
    \label{noise:fig}}
\end{figure}

In the previous example, we have assumed that the AFM is able to measure with a great accuracy the height field $h$. However, AFM resolution is limited in practice, below the nanometer scale~\citep{aliano2012afm}. To validate our approach in a more realistic context, we add a random noise $\xi(\mathbf{r})$ to $h(\mathbf{r})$ on the whole ROI before solving the inverse problem. We choose a correlated noise of maximum amplitude $\pm1$~nm, obtained by filtering a white noise by a spatial Gaussian filter with a standard deviation of $2.5~\mu$m. An example is displayed in Figure~\ref{noise:fig}. 

In this case, $c \simeq0.9994$ and $\rho \simeq 0.59$, as above. The reconstructed pressure field remains close to the applied one in spite of the AFM noise. From now, such an AFM noise with the same characteristics is systematically taken into account. 

Since the intensity of the pressure exerted by the T-cell depends on the substrate rigidity~\citep{Ghibaudo2008,Wahl2019,mandal2023wasp}, we also tested a 5-fold weaker longitudinal pressure field such that $F_{\parallel,{\rm tot}} =2$~nN. The qualitative conclusions remain unchanged and the quantitative indicators read  $c \simeq 0.97$ and $\rho \simeq 0.26$.

\subsubsection{More realistic T-cell synapses}

Even though it retains some axial symmetry character on average, a real immunological synapse is not perfectly circular, nor does it exert a perfectly axisymmetric pressure field~\citep{Basu2017,Bashour2014,Hui2015}. 

For this reason, we first design a non-circular cell support $\mathcal{S}$, the boundary of which is defined in polar coordinates as a constant plus some Fourier modes of random intensity as
\begin{equation}
    r(\theta) = 
    R \left(
    1 + \frac{C_{\text{noise}}}{n_{\text{modes}}}
    \sum_{k=1}^{n_{\text{modes}}}
    \Big[
    a_k \cos(k\theta)
    + b_k \sin(k\theta)
    \Big]
    \right)
\end{equation}
where $R$ is the mean radius, $C_{\text{noise}}=0.1$ controls the amplitude of the shape fluctuations, and $a_k$, $b_k$ are random Fourier coefficients distributed according to a Gaussian distribution with mean 0 and standard deviation 1, for $n_{\text{modes}}= 6$.  We make sure that the net force still vanishes in spite of this modification by subtracting to each pixel the average residual force (i.e. the total force divided by the number of support pixels, that no longer vanishes if the boundary is no longer circular).

\begin{figure}
	\centering
    \includegraphics[height=5cm]{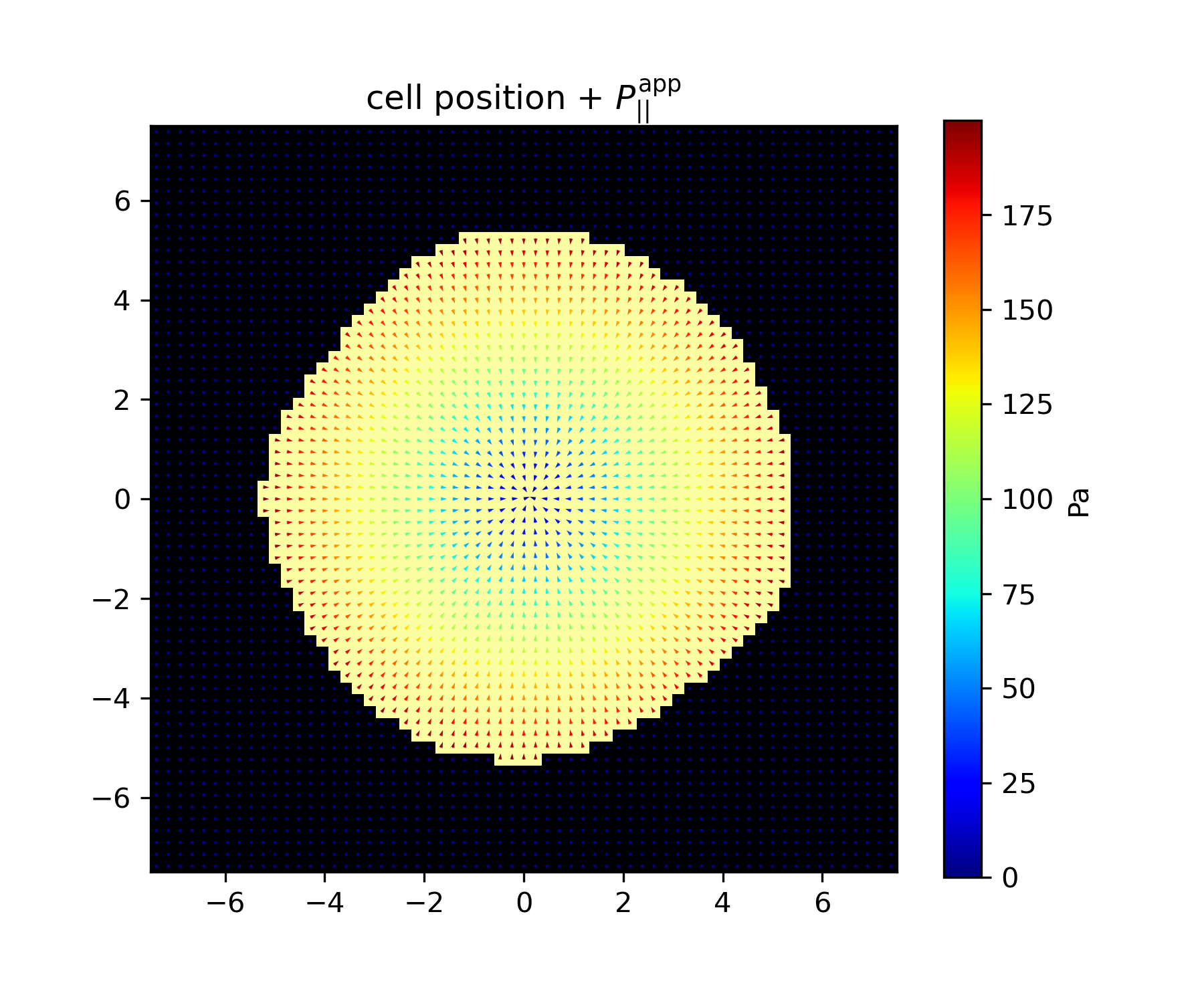}  \hspace{-0.8cm} \includegraphics[height=5cm]{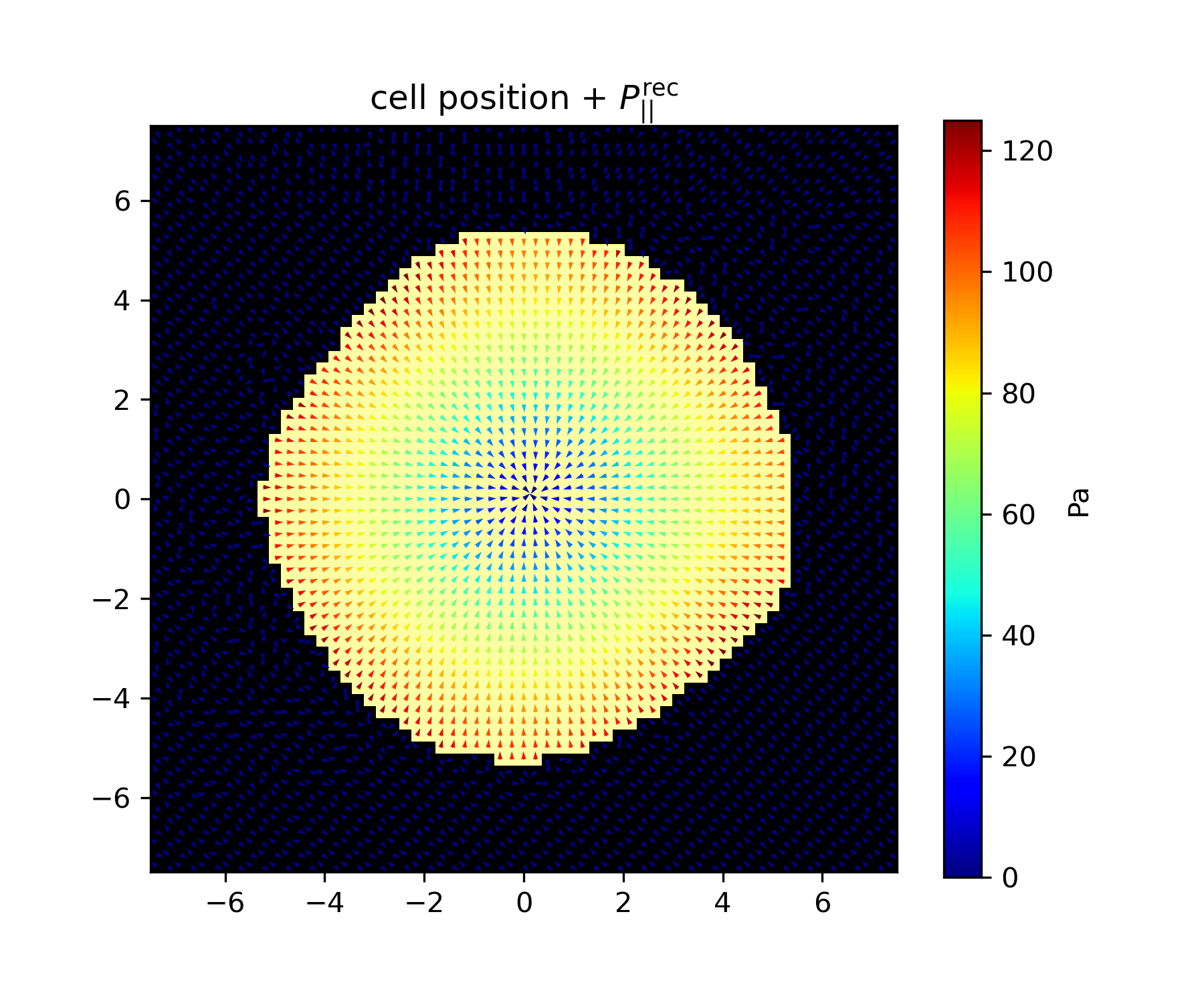}
    \caption{More realistic T-cell synapse with an irregular boundary. Left: applied longitudinal pressure field $P_\parallel(r) \mathbf{e}_r$. Right: reconstructed pressure field. Same parameter values as in Figure~\ref{ideal:synapse}.
    \label{noisy:boundary}}
\end{figure}

The longitudinal applied and reconstructed pressure fields are displayed in Figure~\ref{noisy:boundary}, again showing the good efficacy of our approach, even in the non-symmetric immunological synapse case. Nonetheless, one can notice that some fine features of the applied field are not faithfully restored, notably the width of the annulus of strong pressures (in red) can be locally somewhat misestimated. However, the quantitative indicators read  $\rho \simeq 0.60$ and $c \simeq 0.997$, again comparable to the previous ones.

\begin{figure}
	\centering
    \includegraphics[height=5.45cm]{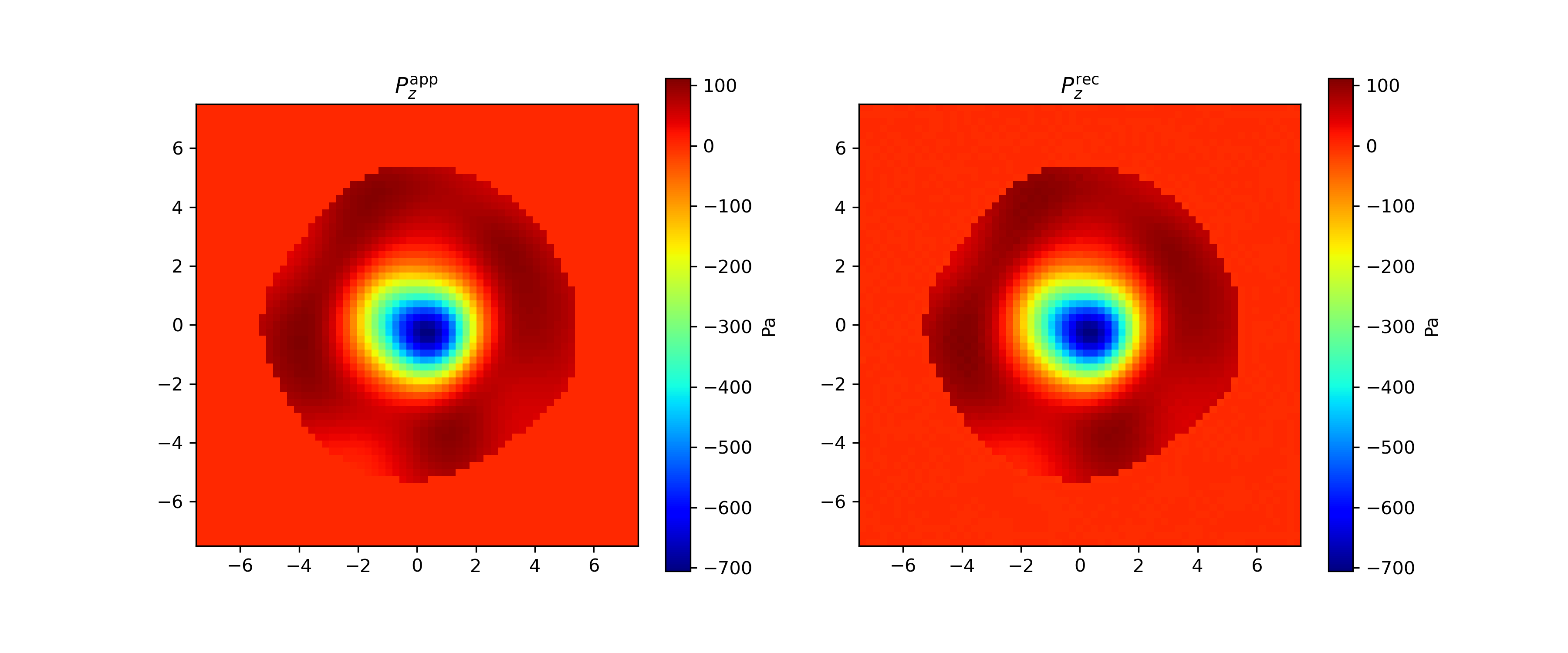} \\
    \includegraphics[height=5cm]{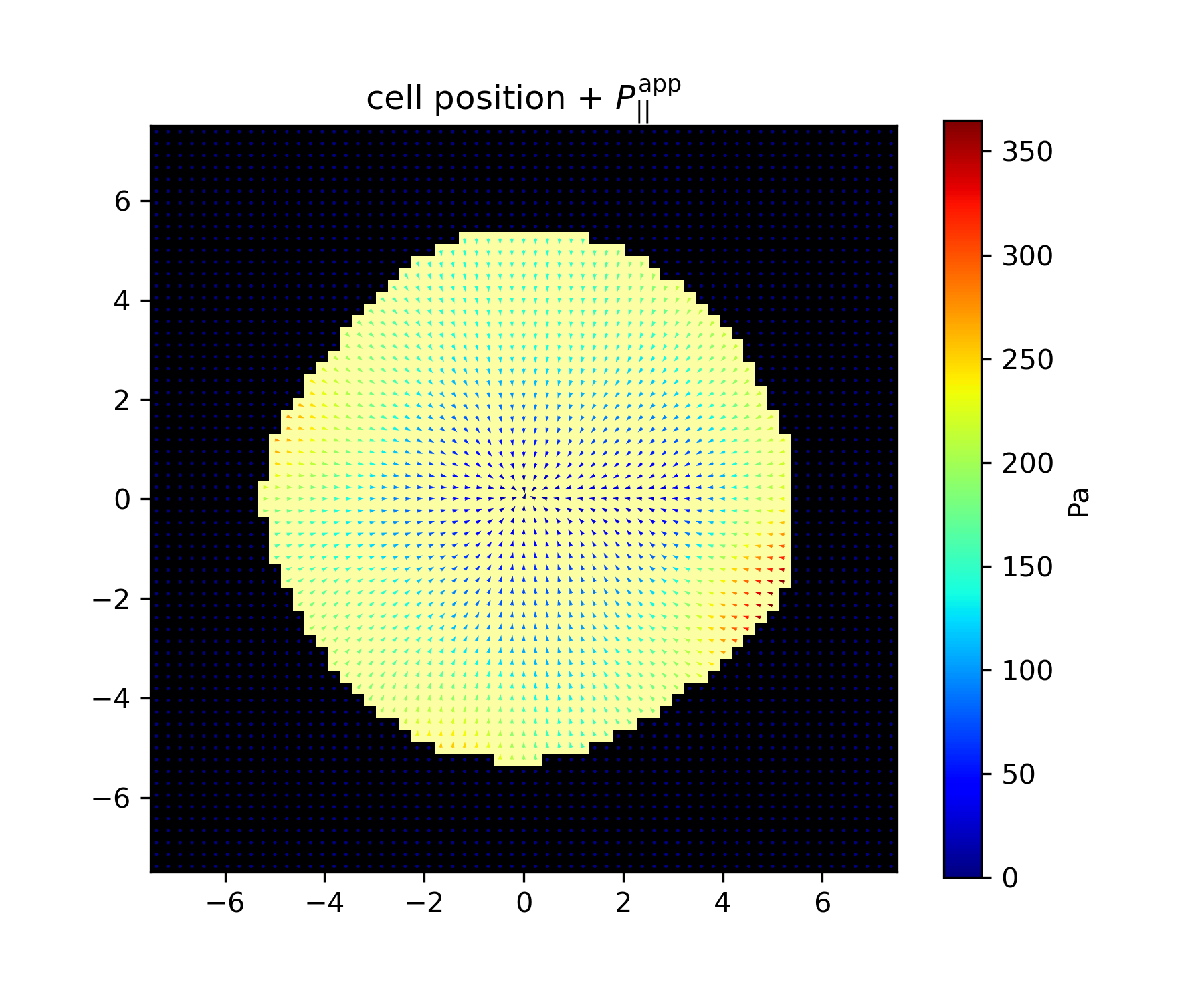}  \hspace{-0.8cm} 
    \includegraphics[height=5cm]{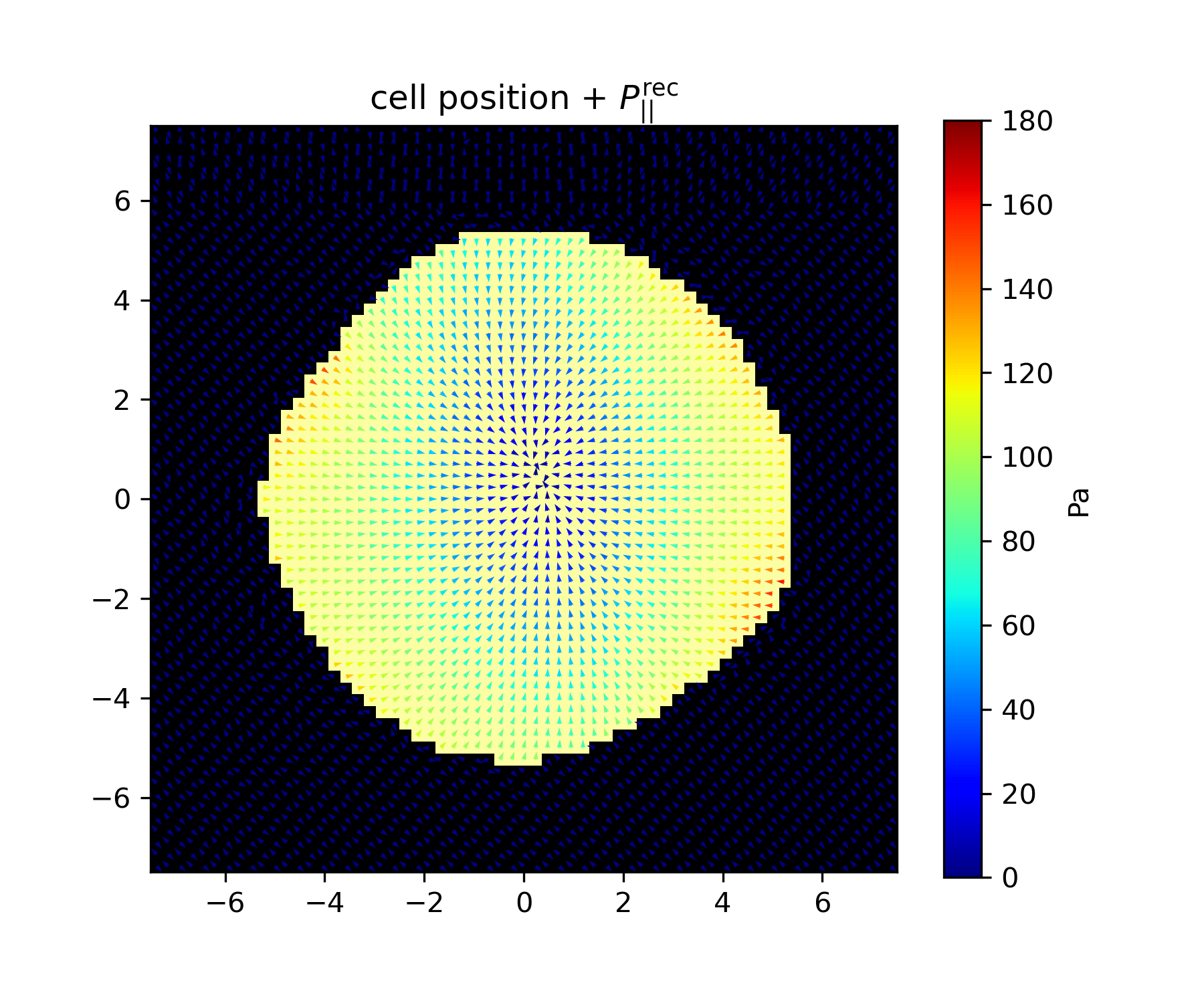}
    \caption{T-cell synapse with an irregular boundary and a non-axisymmetric pressure field. Top-left: applied transverse pressure field $P_z(r)$. Top-right: reconstructed transverse pressure field. 
    Bottem-left: applied longitudinal pressure field $P_\parallel(r)$. Bottom-right: reconstructed longitudinal pressure field. Same parameters as in Figure~\ref{ideal:synapse}.
    \label{noisy:force:fig}}
\end{figure}

In addition to this, we now add some noise on the force field itself to take into account the fact that the pressure exerted by the cell is nor uniform at the single cell level~\citep{Hui2015}. This random noise is generated as follows: a white-noise field is first created over the ROI, for both $x$ and $y$ components of the force field, then convolved with a spatial Gaussian filter with a width of $2.5~\mu$m, according to \citep{huse2020microparticle,Hui2015}. The resulting noise is scaled by an amplitude of $P_\parallel(r)$ (resp. $|P_z(r)|$), meaning that the fluctuations range from $-P_\parallel(r)$ to $P_\parallel(r)$ (resp. $-|P_z(r)|$ to $|P_z(r)|$) relatively to the local force field. The average residual force (i.e. the total force divided by the number of points) is again subtracted from each force component, to ensure that the sum of all forces is still zero. Figure~\ref{noisy:force:fig} illustrates the corresponding pressure patterns (see also Appendix~\ref{noiseExemple} for a different representation of a different example). Again, the transverse pressure field is satisfactorily reconstructed whereas the longitudinal one is shrunk by a factor $\rho \simeq 0.55$; $c \simeq 0.94$ is again close to 1, indicating that the non-homogeneous longitudinal pressure field is correctly reconstructed up to the scaling factor $\rho$ (see Discussion).

We ensured that varying the random seed, $5$ to $10$ times for each situation, had no significant impact on the accuracy of the results.

\section{Conclusion and discussion}

An adherent cell such as a T-cell forming a synapse exerts a 3D pressure field on a substrate, where transverse and longitudinal components have been observed to have the same order of magnitude. After deriving the deformation of an elastic membrane undergoing a 3D pressure field exerted close to its center by such a cell at the level of its immunological synapse, we have demonstrated that it is possible to reconstruct this 3D pressure field $\mathbf{P}(\mathbf{r})$ from the measurement of the sole height $h(\mathbf{r})$ of the membrane, a scalar field. To circumvent the fact that the ensuing system of linear equation is either overdetermined or underdetermined depending on the size of the ROI, we have opted for a formulation in terms of an optimization problem, in the spirit of Reference~\citep{huse2020microparticle}.

As compared to this anterior work~\citep{huse2020microparticle}, where the elasticity problem is solved numerically through a finite-element formulation of the (linear) elastic energy introduced as an additional term in the cost function, we analytically solve the elastic problem {\it a priori}. Even thought it is only an approximation, we have worked at the linear order to be able to benefit from the superposition principle of the linear theory of elasticity. We shall return to this point below. 

Even though we have used the Young modulus of Formvar in our examples because Formvar is the polymer used to make the membrane in the experiments of Reference~\citep{labernadie2014protrusion}, we have been led to use a significantly lower value of the residual bulk tension $\tau_0$ to be able to recover the order of magnitude of the longitudinal pressure field from the measurement of the height field, which indicates that Formvar is probably not the appropriate material for the applicability of our approach to the case of T-cells. T-cells are small cells exerting a moderate force at the immunological synapse level; for larger cells exerting significantly stronger forces, a material with higher residual tension might be appropriate. For example, for a cell of radius $R=25$~$\mu$m~\citep{Li2015} exerting total longitudinal and transverse forces of $300$~nN~\citep{Caille2002}, modeling a larger adherent cell such as a fibroblast, we find $\rho\simeq 0.53$ with a membrane of residual bulk tension $\tau_0=10^3$~Pa (in the axisymmetric case). 

Using a larger residual tension leads to a less favorable ratio $\rho$ of the reconstructed-to-applied longitudinal pressures. For example, we have reproduced the calculations of Section~\ref{illustrative:ex} with a 10-fold stronger tension ($\tau_0=10^3$~Pa) and observed that the value $\rho$ decreased from $\simeq 0.6$ to $\simeq 0.3$ to $0.4$ ($c$ remaining very close to 1). The reason for this shrinking of the reconstructed longitudinal pressure comes from the quadratic form $Q[\mathbf{P}]$ defined in Equation~\eqref{Q:def} that we aim to minimize. Once the angles between the nearest neighbor forces $\mathbf{P}_\parallel^i$ and $\mathbf{P}_\parallel^j$ has been optimized, minimizing $Q$ further is possible by reducing the intensity of the $\mathbf{P}_\parallel^i$. Since the effect of the longitudinal pressure on the height field $h$ is only a correction to the main effect of the transverse one $P_z$, shrinking the $\mathbf{P}_\parallel^i$ can be compensated by a slight modulation of the $P_z^i$. The same effect has been observed in Reference~\citep{huse2020microparticle}, where a ratio $\rho \simeq 0.6$ was also found. By contrast, $P_z$ is not shrunk because it is the main factor contributing to $h$. 

The fact that the shrinking of $\mathbf{P}_\parallel$ depends on the residual tension of the membrane is explained as follows. Larger tension $\tau_0$ implies smaller gradient of $h$: in Equation~\eqref{h_dl}, the consequence of shrinking $\mathbf{P}_\parallel^i$ is reduced. Weaker gradients of $h$ reduce the sensitivity to $\mathbf{P}_\parallel^i$. 

Another parameter with which one can play is the material Young modulus $E$. Its role is two-fold: it controls the longitudinal displacement $\mathbf{u}_\parallel$ through the characteristic length $\tilde{\mathbf{f}}_\parallel$; the longitudinal displacement is thus inversely proportional to $E$ and reducing $E$ should increase the sensibility of the method to the longitudinal pressure field. In contrast, $E$ enters the inverse Helfrich correlation length $k\propto E^{-1/2}$. Reducing $E$ increases $k$ (thus reduces the correlation length) and reduces the range of the Bessel functions in $u_z$. It results that the deformation profile $h$ is more peaked around the origin and is thus less sensible to the longitudinal deformation field $\mathbf{u}_\parallel$ away from the cell support. This reduces in turn the sensibility of the method to the longitudinal pressure field. Consequently, an optimal choice of $E$ must be found to balance
these contradictory trends. As an example, for total forces of 10~nN and a residual bulk tension $\tau_0=1000 Pa$, a ten-fold weaker Young's modulus $E=2.3 \times 10^8$~Pa increases again the ratio $\rho$ to about 0.52 (see Appendix~\ref{noiseExemple}). These considerations show that the elastic membrane mechanical parameters ought to be optimized by our method prior to the experiments.

\begin{figure}
	\centering
    \includegraphics[height=5cm]{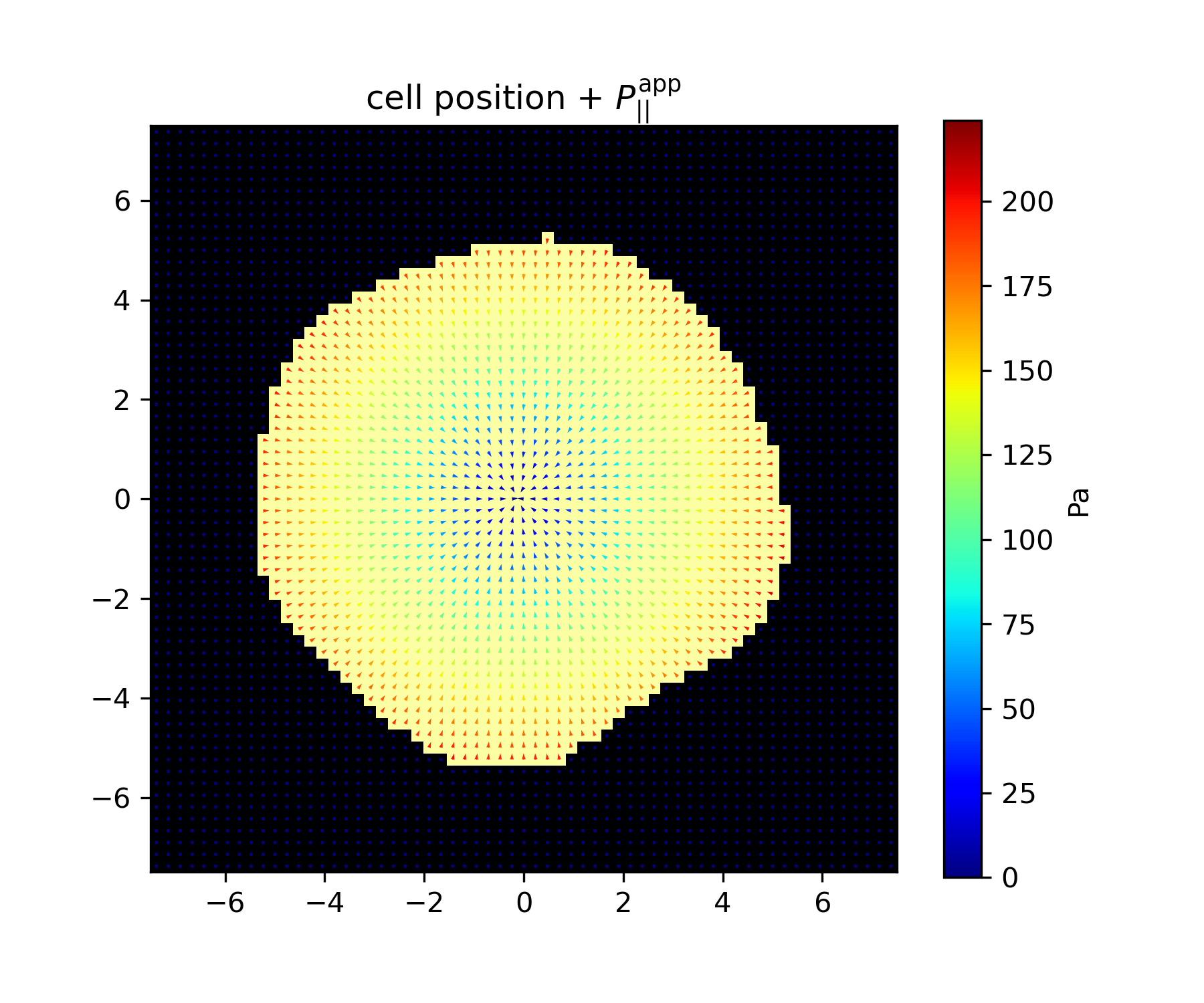}  \hspace{-0.8cm} \includegraphics[height=5cm]{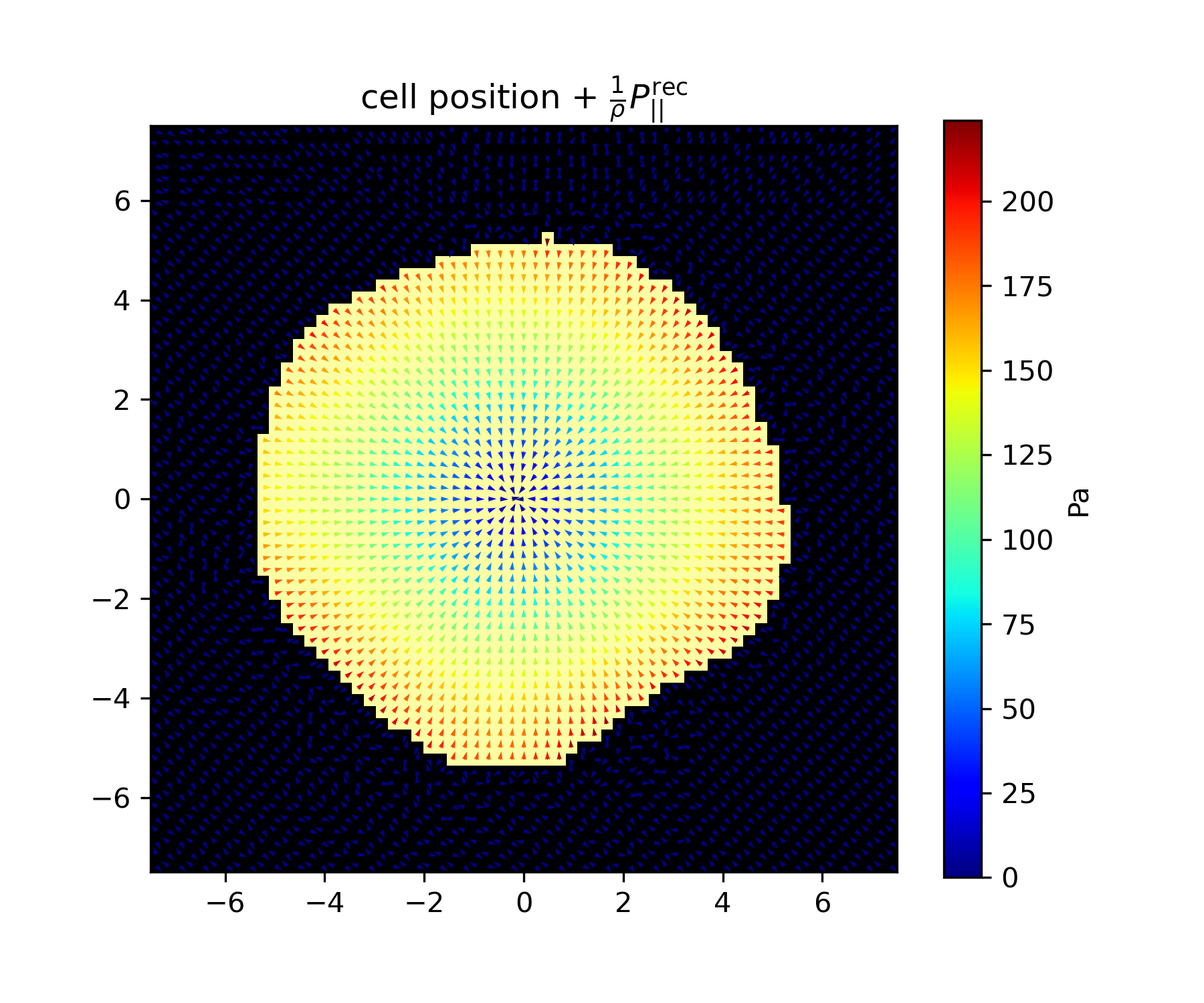}
    \caption{T-cell synapse with an irregular boundary and an axisymmetric pressure field. Left: applied longitudinal pressure field $P_\parallel(r) \mathbf{e}_r$. Right: reconstructed pressure field, multiplied by inverse of the scaling factor $\rho \simeq 0.35$. Parameter values are the same as in Figure~\ref{ideal:synapse}, except that the  residual bulk tension was set to $\tau_0=1000$~Pa.
    \label{asymetrical:rho_factor}}
\end{figure}

Even though this shrinking of the longitudinal field is at first sight a weakness of our approach, we are able to anticipate this effect by the numerical calculation of $\rho$. Up to this effect, the main features of the pressure field are preserved so that the original longitudinal pressure field can essentially be recovered by multiplying the reconstructed one by a global scaling factor $1/\rho$, as shown in Figure~\ref{asymetrical:rho_factor}. The value of $\rho$ can easily be benchmarked by our approach for a given set of material parameters and forces, as long as $\rho$ is not too small. The shrinking of $\mathbf{P}_\parallel$ can thus be compensated for in a practical way. 

To finish with, we need to discuss the small-gradient approximation used to calculate the propagator $u_z$ in Section~\ref{transverse:prop}. As discussed above, the transverse deformation $h$ required to reconstruct the pressure field has to be large enough. In practice, it has the same order of magnitude in our examples as the membrane diameter $2a$, beyond the linear regime~\citep{timoshenko1959theory}. It ensues that in the experimental context, the transverse pressure field reconstructed from the measurement of $h$ by AFM will only be a rough approximation of the true one, even though it will retain its principal features and correct orders of magnitude. Going beyond this small-gradient approximation would require more sophisticated analytical approaches or numerical ones, beyond the scope of the present proof-of-concept study. 

By contrast, the longitudinal pressure field is reconstructed, through Equation~\eqref{h:vs:u2}, from the gradient of $h$, derived from the measured height field $h$ independently on the propagator $u_z$. Consequently, its reconstruction should not suffer significantly from the relative validity of the small gradient of $h$ approximation. To reconstruct the whole 3D pressure field $\mathbf{P}$ on the ROI, an option would then be to minimize the cost function $\Gamma$ numerically, for example by using a gradient descent or a simulated annealing algorithm, in the subspace of $\mathbb{R}^{3N}$ defined by the $N+3$ constraints~\eqref{h_g2} and \eqref{Lagrange:mu}, in which the membrane deformation $u_z$ would be calculated numerically at each iteration. This question ought to be explored in future studies. 

\subsection*{Data availability} 

Data will be made available on request.

\subsection*{Acknowledgements} 

This work was supported by the 80Prime program coordinated by the CNRS MITI. We are also indebted to Renaud Poincloux for his insightful advice on Protrusion Force Microscopy and to Fabien Alet for his valuable contributions on numerical inversion of large matrices.

\bigskip

\appendix

\section*{Appendices}
\section{Deflection of a circular membrane under tension by a transverse point loading}
\label{app0}

The deflection along the $z$ direction of a circular membrane of radius $a$, for a transverse point loading $P_z = F_z\delta(\mathbf r)$ at its center $O$ at $\mathbf r = \mathbf 0$, parallel to the membrane axis of symmetry (Figure~\ref{cartoon}, Bottom), satisfies, for small gradients $\|
\boldsymbol{\nabla} u_z\| \ll 1 $, the ordinary differential equation~\citep{poilane2000analysis,timoshenko1959theory}: 
\begin{equation}
	\frac{\partial^2 \psi}{\partial r^2} + \frac{1}{r}\frac{\partial \psi}{\partial r} - \frac{1}{r^2} \psi = -\frac{1}{\kappa} \frac{F_z}{2\pi r}
	\label{eq_psi_no_tension}
\end{equation} where $\psi(\mathbf r)$ is the small angle between the normal to the deflection and the axis of symmetry of the membrane $(Oz)$, defined for small gradients as:
\begin{equation}
	\psi(\mathbf r) = - \frac{\partial u_z}{\partial r}
\end{equation} where $u_z(\mathbf r)$ is the deflection of the membrane, $r = \|\mathbf r\| = \sqrt{x^2+y^2}$ is the distance to the center $O$, 
$\kappa =\frac{E e^3}{12(1 - \nu^2)} $ is the bending rigidity with $E$ the Young modulus, $e$ the membrane thickness and $\nu$ the Poisson ratio~\citep{landau2012theory}.  
The term on the right-hand-side, $\frac{F_z}{2\pi r}$, is the shearing force per unit length. 

A key factor influencing the membrane deformation is the residual tension within the membrane, inherent to the material fabrication process~\citep{poilane2000analysis}. When the membrane undergoes a uniform frame tension $\tau > 0$, the shearing force becomes $\frac{F_z}{2\pi r} - \tau \psi (\mathbf r)$ and the equation \eqref{eq_psi_no_tension} becomes the Bessel equation \citep{abramowitz1968handbook}:
\begin{equation}
	\frac{\partial^2 \psi}{\partial r^2} + \frac{1}{r}\frac{\partial \psi}{\partial r} -\left( \frac{k^2}{a^2} + \frac{1}{r^2} \right) \psi = -\frac{1}{\kappa} \frac{F_z}{2\pi r}
	\label{uz}
\end{equation} 
where $k = a\sqrt{\frac{ \tau}{\kappa}}$ is the dimensionless inverse Helfrich correlation length~\citep{hong1990measuring}. This differential equation can also be derived from the minimization of the Helfrich free energy of a membrane under tension in the small gradient approximation~\citep{Safran1994,Julicher1994}.

If the membrane is clamped at its edges, \textit{i.e.} $u_z(a) = 0$ and $ \frac{\partial u_z}{\partial r}|_{r=a} = 0$, as assumed here, the general solution for $u_z = - \int \psi {\rm d} r $ is \citep{hong1990measuring}: 

\begin{align}
	u_z (r) &= \frac{F_za^2}{2\pi  \kappa k^2} \left\{  \frac{(K_1(k)-\frac{1}{k})}{I_1(k)}\left(I_0(k) - I_0 \left( \frac{kr}{a}\right) \right) + \left(K_0(k) - K_0 \left( \frac{kr}{a} \right) \right) - \text{ln} \left( \frac{r}{a} \right)  \right\}
	\label{u_z_app}
\end{align} where $I_{n}$ and $K_{n}$ are respectively modified Bessel functions of first and second kind of order $n$~\citep{abramowitz1968handbook}. At $r=0$, $K_0(r)$ diverges logarithmically but the divergence is canceled by the logarithm and 
\begin{align}
	u_z(0) &=  \frac{F_z a^2}{16\pi  \kappa }  g(k) 
	\label{uz0}
\end{align}  where
\begin{align}
	g(k)&= \frac{8}{k^2}\left\{  \left(K_1(k)-\frac{1}{k}\right)  \frac{(I_0(k) - 1)}{I_1(k)} + \left(K_0(k)+\text{ln} \left(\frac{k}{2} \right)+ \gamma \right)  \right\}
	\label{g(k)}
\end{align} $\gamma$ is Euler's constant and $g(k)$ represents the contribution of the tension to the vertical deflection. Indeed when $\tau =0$, $g(0) = 1$ and one recovers the solution at $r=0$ for a tensionless membrane $u_z(0) = \frac{F_z a^2}{16\pi \kappa}$  \citep{timoshenko1959theory}. In the other limit $k \gg 1$, in practice as soon as $k\geq 5$, $g(k) \simeq 8 \ln(k)/k^2$, so that  
\begin{align}
	u_z(0) &\simeq  \frac{F_z}{2\pi  \tau }  \ln k 
	\label{uz0bis}
\end{align} 
\section{Boundary condition under longitudinal loading}
\label{appA}
 When a local longitudinal force $\mathbf{F}_\parallel$ is applied at the center $O$ of a circular membrane of radius $a$, the biharmonic function $\phi$ (i.e. respecting $\Delta \Delta\phi(\mathbf r) = 0$) in Equation~\eqref{psi:phi} can be chosen so that the displacement $\mathbf{u}_\parallel(\mathbf{r})$ vanishes when $\| \mathbf{r} \|=a$. Indeed, in polar coordinates, biharmonic functions $\phi$ leading to regular displacements at the origin are a linear combination of the functions~\citep{michell1899direct} $\phi_{n,p}(r,\theta)= \cos(n\theta)\, r^p$ with $n\geq 0$ and $p=n$ or $p=n+2$, and the same functions where the $\cos$ is replaced by a $\sin$. 
 
 By writing in polar coordinates the displacement $\mathbf{u}_\parallel(r,\theta)$ associated with the first functions $\phi_{n,p}$, one finds that the following function
 \begin{equation}
 \phi(r,\theta)=\frac{a^2}{4\pi}\left\{
 4\frac{\nu-1}{(\nu-3)^2} \left(\frac{r}{a}\right)^2 + \left[-\frac{\nu+1}{6(\nu-3)} \cos(2\theta) +\frac1{24}\cos(4\theta)\right] \left(\frac{r}{a}\right)^4
 \right\}
 \end{equation}
 satisfies the fixed boundary condition. The displacement given in the main text derives from this biharmonic function.

\section{Force applied close to the membrane center}
\label{appB}

We first show that if the longitudinal force $\mathbf{F}_\parallel$ is applied close to the center, at a point $\mathbf{r}_0$ such that $\| \mathbf{r}_0 \|/ a \ll 1$, then the displacement Equation~\eqref{u_r_vect} is unchanged at leading order, up to a shift: 
\begin{equation}
\mathbf{u}_\parallel(\br) = \mathbf{G}_\parallel(\br - \br_0) \label{propa}
\end{equation}
where we have introduced the propagator (Equation~\eqref{u_r_vect})
\begin{eqnarray}
	\mathbf{G}_\parallel(\br) & = & -\frac{(\nu+1)^2(3\nu-1)}{8\pi(\nu-3)} \left[\left(\frac{r}{a}\right)^2-1\right] \frac{\tilde{\mathbf{f}}_\parallel\cdot\mathbf{r}}{r^2} \mathbf{r} \nonumber \\ 
	& + & \frac{(\nu+1)^2(\nu+1)}{8\pi(\nu-3)} \left[\left(\frac{r}{a}\right)^2-1\right] \frac{\tilde{\mathbf{f}}_\parallel \times \mathbf{r}}{r}
	+ \frac{(\nu-3)(\nu+1)}{4\pi}   \ln\left(\frac{r}{a}\right) \tilde{\mathbf{f}}_\parallel  \nonumber \\
	\label{propagatorG}
\end{eqnarray} 

Indeed, applying the force at $\br_0$ amonts to shift the whole system by $-\br_0$ and to apply the force at the origin $O$. In polar coordinates, the boundary circle now writes  $r(\theta) = a - r_0 \cos(\theta-\theta_0)$ at first order in $\frac{r_0}{a}$, where $\theta_0$ is the polar angle of $\br_0$. Since the force is applied at the origin, we must simply check that $\mathbf{u}_\para(\br)$ given in Equation~\eqref{u_r_vect} still vanishes on the new boundary $r(\theta)$. We  demonstrate that this is true at dominant order in the two infinitesimals $\frac{\|\tilde{\mathbf{f}}_\para\|}{a}$ and $\frac{r_0}{a}$: Replacing $r$ by $r(\theta)$ in Equation~\eqref{propagatorG}, and expanding again the expression at order one in $\frac{r_0}{a}$, one is only left with terms of order 2, in $\frac{r_0}{a}\|\tilde{\mathbf{f}}_\parallel\|$. Shifting again the system by $+\br_0$ to bring back the point of application of $\mathbf{F}_\para$ to its original position eventually leads to relation~\eqref{propa}.

The same approach can be followed to show that the propagator $G_z$ defined through $u_z$ in Equation~\eqref{uz} can  be used in a similar way when the point of application of the transverse force $\mathbf{F}_z$ is at a distance $r_0 \ll a$ from the membrane center.

\section{Scalar product between random unitary vectors of $\mathbb{R}^d$}
\label{scalar:prod}
 Let $\mathbf{u}$ be a fixed unitary vector of $\mathbb{R}^d$, chosen as a base vector, $\mathbf{u} = \mathbf{e}_1$, without loss of generality. Let $\mathbf{x}$ be another vector of $\mathbb{R}^d$, not necessarily unitary, the coordinated of which are $d$ independent Gaussian random variables of zero mean and standard deviation $s$, and $\mathbf{u'} = \mathbf{x}/\| \mathbf{x} \|$. We demonstrate that $\mathbf{u}\cdot \mathbf{u'} \sim 1/\sqrt{d}$. 
 
 Indeed $\langle \| \mathbf{x} \|^2 \rangle = d \, s^2$, and when $d$ is large, $\| \mathbf{x} \|^2$ is peaked around this mean value so that $\| \mathbf{x} \|^2 \simeq d \, s^2$. Consequently, $\mathbf{u}\cdot \mathbf{u'} \simeq \mathbf{x}\cdot \mathbf{e}_1 /(s \sqrt{d})$. Now, $\mathbf{x}\cdot \mathbf{e}_1$ being a coordinate of $\mathbf{x}$, it is a Gaussian random variables of zero mean and standard deviation $s$. Thus the standard deviation of $\mathbf{u}\cdot \mathbf{u'}$, setting its order of magnitude, is $s/(s \sqrt{d}) = 1/\sqrt{d}$. This prediction has been successfully tested numerically with excellent accuracy.

\section{Noise on the longitudinal pressure field}
\label{noiseExemple}

\begin{figure}
	\centering
	\includegraphics[height=8cm]{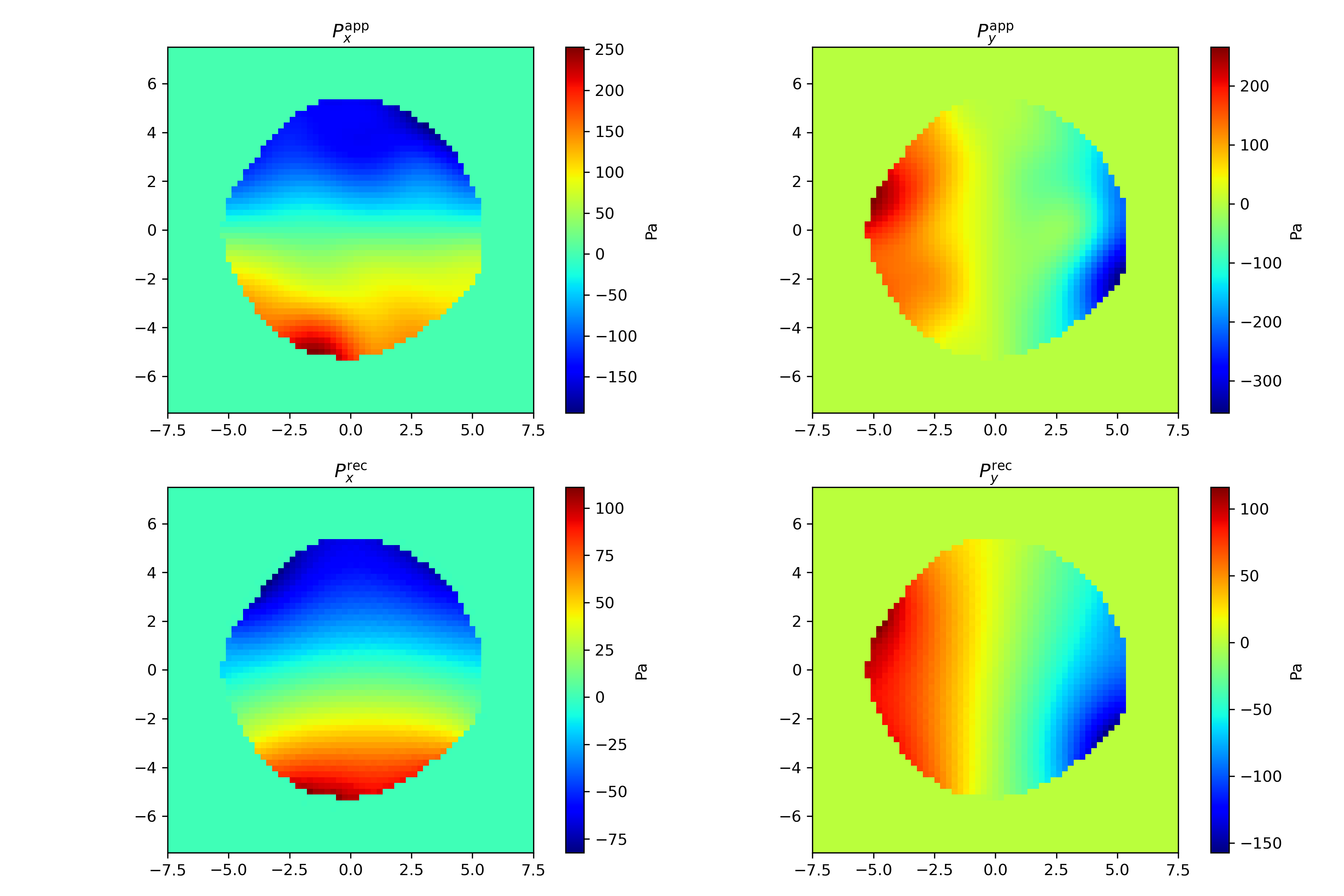}
    \caption{T-cell synapse with an irregular boundary and an non-axisymmetric pressure field. Top-left: applied longitudinal pressure field $P^{\mathrm{app}}_x$. Top-right: applied longitudinal pressure field $P^{\mathrm{app}}_y$. Bottom-left: reconstructed longitudinal pressure field $P^{\mathrm{rec}}_x$ Bottom-right: reconstructed longitudinal pressure field $P^{\mathrm{rec}}_y$. Same parameter values as in Figure~\ref{ideal:synapse} except that the  residual bulk tension was set to $\tau_0=1000$~Pa and the Young modulus to $E = 2.3\times 10^8$~Pa.
    \label{asymetrical:x_y_noise}}
\end{figure}

To derive the non-axisymmetric pressure field
and simulate a more realistic T-cell synapse, we add an independent noise to both the $x$ and $y$ components of $\mathbf{P}_\parallel = P_x \mathbf{e}_x+P_y \mathbf{e}_y$. The noise is implemented as a white noise convolved with a Gaussian filter of width 2.5 µm, and it is scaled relatively to the local force field, allowing fluctuations of up to $\pm P_\parallel (r)$, as shown in Figure~\ref{asymetrical:x_y_noise}.

The reconstructed field is more regular than the applied one because of the quadratic form $Q$ in the cost function. 

The same procedure is applied to $P_z$. 


\bibliographystyle{unsrt}
\bibliography{biblio}

@article{Caille2002,
	title = {Contribution of the nucleus to the mechanical properties of endothelial cells},
	volume = {35},
	copyright = {https://www.elsevier.com/tdm/userlicense/1.0/},
	issn = {00219290},
	url = {https://linkinghub.elsevier.com/retrieve/pii/S0021929001002019},
	doi = {10.1016/S0021-9290(01)00201-9},
	language = {en},
	number = {2},
	urldate = {2025-08-13},
	journal = {Journal of Biomechanics},
	author = {Caille, N. and Thoumine, O. and Tardy, Y. and Meister, J.-J.},
	year = {2002},
	pages = {177--187},
}

@article{Li2015,
	title = {Moving {Cell} {Boundaries} {Drive} {Nuclear} {Shaping} during {Cell} {Spreading}},
	volume = {109},
	issn = {00063495},
	url = {https://linkinghub.elsevier.com/retrieve/pii/S0006349515007109},
	doi = {10.1016/j.bpj.2015.07.006},
	language = {en},
	number = {4},
	urldate = {2025-08-13},
	journal = {Biophysical Journal},
	author = {Li, Y. and Lovett, D. and Zhang, Q. and Neelam, S. and Kuchibhotla, R.A. and Zhu, R. and Gundersen, G.G. and Lele, T.P.. and Dickinson, R.B.},
	year = {2015},
	pages = {670--686},
}

@article{Huse2025,
	title = {Mechanoregulation of lymphocyte cytotoxicity},
	issn = {1474-1733, 1474-1741},
	url = {https://www.nature.com/articles/s41577-025-01173-2},
	doi = {10.1038/s41577-025-01173-2},
	language = {en},
	urldate = {2025-08-11},
	journal = {Nat Rev Immunol},
	author = {Huse, M.},
	year = {2025},
}

@article{Dillard2014,
	title = {Ligand-{Mediated} {Friction} {Determines} {Morphodynamics} of {Spreading} {T} {Cells}},
	volume = {107},
	issn = {00063495},
	url = {https://linkinghub.elsevier.com/retrieve/pii/S0006349514011308},
	doi = {10.1016/j.bpj.2014.10.044},
	language = {en},
	number = {11},
	urldate = {2025-08-13},
	journal = {Biophysical Journal},
	author = {Dillard, P. and Varma, R. and Sengupta, K. and Limozin, L.},
	year = {2014},
	pages = {2629--2638},
}

@book{NumericalRecipes,
  asin = {0521880688},
  author = {Press, W.H. and Teukolsky, S.A. and Vetterling, W.T. and Flannery, B.P.},
  dewey = {518.0285},
  ean = {9780521880688},
  edition = 3,
  isbn = {0521880688},
  publisher = {Cambridge University Press},
  title = {Numerical Recipes 3rd Edition: The Art of Scientific Computing},
  url = {http://www.amazon.com/Numerical-Recipes-3rd-Scientific-Computing/dp/0521880688/ref=sr_1_1?ie=UTF8&s=books&qid=1280322496&sr=8-1},
  year = 2007
}

@book{Safran1994,
author = {Safran, S.A.},
address = {Reading (Mass.)},
language = {eng},
publisher = {Addison-Wesley},
series = {Frontiers in physics 90},
title = {Statistical Thermodynamics of Surfaces, Interfaces, and Membranes},
year = {1994},
isbn = {0201626330},
}

@article{Julicher1994,
	title = {Shape equations for axisymmetric vesicles: {A} clarification},
	volume = {49},
	copyright = {http://link.aps.org/licenses/aps-default-license},
	issn = {1063-651X, 1095-3787},
	shorttitle = {Shape equations for axisymmetric vesicles},
	url = {https://link.aps.org/doi/10.1103/PhysRevE.49.4728},
	doi = {10.1103/PhysRevE.49.4728},
	language = {en},
	number = {5},
	urldate = {2025-08-17},
	journal = {Phys. Rev. E},
	author = {Jülicher, F. and Seifert, U.},
	year = {1994},
	pages = {4728--4731},
}

@article{Chen2017,
author = {Chen, Y. and Ju, L. and Rushdi, M. and Ge, C. and Zhu, C.},
title = {Receptor-mediated cell mechanosensing},
journal = {Molecular Biology of the Cell},
volume = {28},
number = {23},
pages = {3134-3155},
year = {2017},
doi = {10.1091/mbc.e17-04-0228},
    note ={PMID: 28954860},
URL = { 
        https://doi.org/10.1091/mbc.e17-04-0228
}
}

@article{Hui2015,
	title = {Cytoskeletal forces during signaling activation in {Jurkat} {T}-cells},
	volume = {26},
	issn = {1059-1524, 1939-4586},
	url = {https://www.molbiolcell.org/doi/10.1091/mbc.E14-03-0830},
	doi = {10.1091/mbc.E14-03-0830},
	language = {en},
	number = {4},
	urldate = {2025-08-13},
	journal = {MBoC},
	author = {Hui, K.L. and Balagopalan, L/ and Samelson, L.E. and Upadhyaya, A.},
	editor = {Wang, Yu-Li},
	year = {2015},
	pages = {685--695},
}

@article{Blumenthal2020,
	title = {Multiple actin networks coordinate mechanotransduction at the immunological synapse},
	volume = {219},
	issn = {0021-9525, 1540-8140},
	url = {https://rupress.org/jcb/article/219/2/e201911058/133650/Multiple-actin-networks-coordinate},
	doi = {10.1083/jcb.201911058},
	language = {en},
	number = {2},
	urldate = {2025-08-13},
	journal = {Journal of Cell Biology},
	author = {Blumenthal, D. and Burkhardt, J.K.},
	year = {2020},
	pages = {e201911058},
}

@article{Tamzalit2019,
	title = {Interfacial actin protrusions mechanically enhance killing by cytotoxic {T} cells},
	volume = {4},
	issn = {2470-9468},
	url = {https://www.science.org/doi/10.1126/sciimmunol.aav5445},
	doi = {10.1126/sciimmunol.aav5445},
	language = {en},
	number = {33},
	urldate = {2025-08-13},
	journal = {Sci. Immunol.},
	author = {Tamzalit, F. and Wang, M.S. and Jin, W. and Tello-Lafoz, M. and Boyko, V. and Heddleston, J.M. and Black, C.T. and Kam, L.C. and Huse, M.},
	year = {2019},
	pages = {eaav5445},
}

@article{Wahl2019,
	title = {Biphasic mechanosensitivity of {T} cell receptor-mediated spreading of lymphocytes},
	volume = {116},
	issn = {0027-8424, 1091-6490},
	url = {https://pnas.org/doi/full/10.1073/pnas.1811516116},
	doi = {10.1073/pnas.1811516116},
	language = {en},
	number = {13},
	urldate = {2025-09-11},
	journal = {Proc. Natl. Acad. Sci. U.S.A.},
	author = {Wahl, A. and Dinet, A. and Dillard, P. and Nassereddine, A. and Puech, P.-H. and Limozin, L. and Sengupta, K.},
	year = {2019},
	pages = {5908--5913},
}

@article{Roy2018,
	title = {The {Actin} {Cytoskeleton}: {A} {Mechanical} {Intermediate} for {Signal} {Integration} at the {Immunological} {Synapse}},
	volume = {6},
	issn = {2296-634X},
	shorttitle = {The {Actin} {Cytoskeleton}},
	url = {https://www.frontiersin.org/article/10.3389/fcell.2018.00116/full},
	doi = {10.3389/fcell.2018.00116},
	language = {en},
	urldate = {2025-08-13},
	journal = {Front. Cell Dev. Biol.},
	author = {Roy, Nathan H. and Burkhardt, Janis K.},
	year = {2018},
	pages = {116},
}

@article{Basu2016,
	title = {Cytotoxic {T} {Cells} {Use} {Mechanical} {Force} to {Potentiate} {Target} {Cell} {Killing}},
	volume = {165},
	issn = {00928674},
	url = {https://linkinghub.elsevier.com/retrieve/pii/S0092867416000611},
	doi = {10.1016/j.cell.2016.01.021},
	language = {en},
	number = {1},
	urldate = {2025-08-13},
	journal = {Cell},
	author = {Basu, R. and Whitlock, B.M. and Husson, J. and Le Floc’h, A. and Jin, W. and others},
	year = {2016},
	pages = {100--110},
}

@article{Sawicka2017,
	title = {Micropipette force probe to quantify single-cell force generation: application to {T}-cell activation},
	volume = {28},
	issn = {1059-1524, 1939-4586},
	shorttitle = {Micropipette force probe to quantify single-cell force generation},
	url = {https://www.molbiolcell.org/doi/10.1091/mbc.e17-06-0385},
	doi = {10.1091/mbc.e17-06-0385},
	language = {en},
	number = {23},
	urldate = {2025-08-13},
	journal = {MBoC},
	author = {Sawicka, A. and Babataheri, A. and Dogniaux, S. and Barakat, A.I. and Gonzalez-Rodriguez, D. and Hivroz, C. and Husson, J.},
	editor = {Théry, Manuel},
	year = {2017},
	pages = {3229--3239},
}

@article{Ghibaudo2008,
	title = {Traction forces and rigidity sensing regulate cell functions},
	volume = {4},
	issn = {1744-683X, 1744-6848},
	url = {https://xlink.rsc.org/?DOI=b804103b},
	doi = {10.1039/b804103b},
	language = {en},
	number = {9},
	urldate = {2025-08-13},
	journal = {Soft Matter},
	author = {Ghibaudo, M. and Saez, A. and Trichet, L. and Xayaphoummine, A. and Browaeys, J. and Silberzan, P. and Buguin, A. and Ladoux, B.},
	year = {2008},
	pages = {1836},
}

@article{Bashour2014,
	title = {{CD28} and {CD3} have complementary roles in {T}-cell traction forces},
	volume = {111},
	issn = {0027-8424, 1091-6490},
	url = {https://pnas.org/doi/full/10.1073/pnas.1315606111},
	doi = {10.1073/pnas.1315606111},
	language = {en},
	number = {6},
	urldate = {2025-08-13},
	journal = {Proc. Natl. Acad. Sci. U.S.A.},
	author = {Bashour, K.T. and Gondarenko, A. and Chen, H. and Shen, K. and Liu, XiX.n and Huse, M. and Hone, J.C. and Kam, L.C.},
	year = {2014},
	pages = {2241--2246},
}

@article{Aramesh2021,
	title = {Functionalized {Bead} {Assay} to {Measure} {Three}-dimensional {Traction} {Forces} during {T}-cell {Activation}},
	volume = {21},
	copyright = {https://doi.org/10.15223/policy-029},
	issn = {1530-6984, 1530-6992},
	url = {https://pubs.acs.org/doi/10.1021/acs.nanolett.0c03964},
	doi = {10.1021/acs.nanolett.0c03964},
	language = {en},
	number = {1},
	urldate = {2025-08-13},
	journal = {Nano Lett.},
	author = {Aramesh, M. and Mergenthal, S. and Issler, M. and Plochberger, B. and Weber, F. and others},
	year = {2021},
	pages = {507--514},
}

@article{CallanJones2008,
	title = {Viscous-{Fingering}-{Like} {Instability} of {Cell} {Fragments}},
	volume = {100},
	copyright = {http://link.aps.org/licenses/aps-default-license},
	issn = {0031-9007, 1079-7114},
	url = {https://link.aps.org/doi/10.1103/PhysRevLett.100.258106},
	doi = {10.1103/PhysRevLett.100.258106},
	language = {en},
	number = {25},
	urldate = {2025-08-13},
	journal = {Phys. Rev. Lett.},
	author = {Callan-Jones, A.C. and Joanny, J.-F. and Prost, J.},
	year = {2008},
	pages = {258106},
}

@article{Comrie2015,
	title = {F-actin flow drives affinity maturation and spatial organization of {LFA}-1 at the immunological synapse},
	volume = {208},
	issn = {1540-8140, 0021-9525},
	url = {https://rupress.org/jcb/article/208/4/475/38005/F-actin-flow-drives-affinity-maturation-and},
	doi = {10.1083/jcb.201406121},
	language = {en},
	number = {4},
	urldate = {2025-09-11},
	journal = {Journal of Cell Biology},
	author = {Comrie, W.A. and Babich, A. and Burkhardt, J.K.},
	year = {2015},
	pages = {475--491},
}

@article{Basu2017,
	title = {Mechanical {Communication} at the {Immunological} {Synapse}},
	volume = {27},
	issn = {09628924},
	url = {https://linkinghub.elsevier.com/retrieve/pii/S0962892416301763},
	doi = {10.1016/j.tcb.2016.10.005},
	language = {en},
	number = {4},
	urldate = {2025-08-13},
	journal = {Trends in Cell Biology},
	author = {Basu, R. and Huse, M.},
	year = {2017},
	pages = {241--254},
}

@article{Sens2013,
	title = {Rigidity sensing by stochastic sliding friction},
	volume = {104},
	copyright = {http://iopscience.iop.org/info/page/text-and-data-mining},
	issn = {0295-5075, 1286-4854},
	url = {https://iopscience.iop.org/article/10.1209/0295-5075/104/38003},
	doi = {10.1209/0295-5075/104/38003},
	language = {en},
	number = {3},
	urldate = {2025-09-11},
	journal = {EPL},
	author = {Sens, P.},
	year = {2013},
	pages = {38003},
}

@article{Dembo1996,
title = {Imaging the traction stresses exerted by locomoting cells with the elastic substratum method},
journal = {Biophysical Journal},
volume = {70},
number = {4},
pages = {2008-2022},
year = {1996},
issn = {0006-3495},
doi = {https://doi.org/10.1016/S0006-3495(96)79767-9},
url = {https://www.sciencedirect.com/science/article/pii/S0006349596797679},
author = {Dembo, Micah and Oliver, Tim N. and Ishihara, Akira and Jacobson, Ken A.},
}

@article{huse2020microparticle,
  title={Microparticle traction force microscopy reveals subcellular force exertion patterns in immune cell--target interactions},
  author={Vorselen, D. and Wang, Y. and de Jesus, M.M. and Shah, P.K. and Footer, M.J. and Huse, M. and Cai, W. and Theriot, J.A.},
  journal={Nature communications},
  volume={11},
  number={1},
  pages={20},
  year={2020},
  publisher={Nature Publishing Group UK London}
}

@article{labernadie2014protrusion,
  title={Protrusion force microscopy reveals oscillatory force generation and mechanosensing activity of human macrophage podosomes},
  author={Labernadie, A. and Bouissou, A. and Delobelle, P. and Balor, S. and Voituriez, R. and others},
  journal={Nature communications},
  volume={5},
  number={1},
  pages={5343},
  year={2014},
  publisher={Nature Publishing Group UK London}
}

@book{landau2012theory,
  title={Theory of elasticity},
  author={Landau, L.D. and Pitaevskii, L.P. and Kosevich, Arnold M. and Lifshitz, E.M.},
  volume={7},
  year={1970},
}

@book{abramowitz1968handbook,
  title={Handbook of mathematical functions with formulas, graphs, and mathematical tables},
  author={Abramowitz, M. and Stegun, I.A.},
  volume={55},
  year={1968},
  publisher={US Government printing office}
}

@book{timoshenko1959theory,
  title={Theory of plates and shells},
  author={Timoshenko, S. and Woinowsky-Krieger, S.},
  volume={2},
  year={1959},
  publisher={McGraw-hill New York}
}

@article{poilane2000analysis,
  title={Analysis of the mechanical behavior of shape memory polymer membranes by nanoindentation, bulging and point membrane deflection tests},
  author={Poil{\^a}ne, Christophe and Delobelle, P and Lexcellent, C and Hayashi, S and Tobushi, H},
  journal={Thin Solid Films},
  volume={379},
  number={1-2},
  pages={156--165},
  year={2000},
  publisher={Elsevier}
}

@article{hong1990measuring,
  title={Measuring stiffnesses and residual stresses of silicon nitride thin films},
  author={Hong, S and Weihs, Timothy P. and Bravman, John C. and Nix, William D.},
  journal={Journal of Electronic Materials},
  volume={19},
  pages={903--909},
  year={1990},
  publisher={Springer}
}

@article{michell1899direct,
  title={On the direct determination of stress in an elastic solid, with application to the theory of plates},
  author={Michell, John H.},
  journal={Proceedings of the London Mathematical Society},
  volume={1},
  number={1},
  pages={100--124},
  year={1899},
  publisher={Wiley Online Library}
}

@article{mandal2023wasp,
  title={WASP facilitates tumor mechanosensitivity in T lymphocytes},
  author={Mandal, Srishti and Melo, Mariane and Gordiichuk, Pavlo and Acharya, Sayanti and Poh, Yeh-Chuin and Li, Na and Aung, Aereas and Dane, Eric L and Irvine, Darrell J and Kumari, Sudha},
  journal={bioRxiv},
  year={2023}
}

@article{butler2002traction,
  title={Traction fields, moments, and strain energy that cells exert on their surroundings},
  author={Butler, J.P. and Tolic-N{\o}rrelykke, I.M. and Fabry, B. and Fredberg, J.J.},
  journal={American Journal of Physiology-Cell Physiology},
  volume={282},
  number={3},
  pages={C595--C605},
  year={2002},
  publisher={American Physiological Society Bethesda, MD}
}

@article{schwarz2015traction,
  title={Traction force microscopy on soft elastic substrates: A guide to recent computational advances},
  author={Schwarz, U.S. and Soin{\'e}, J.R.D.},
  journal={Biochimica et Biophysica Acta (BBA)-Molecular Cell Research},
  volume={1853},
  number={11},
  pages={3095--3104},
  year={2015},
  publisher={Elsevier}
}

@article{aliano2012afm,
  title={AFM in liquids},
  author={Aliano, A. and Cicero, G. and Nili, H. and Green, N. and others},
  journal={Encyclopedia of nanotechnology. Springer, Dordrecht},
  pages={83--89},
  year={2012}
}

\end{document}